\documentclass[11pt,a4paper]{article}

\usepackage[
a4paper,
left=28mm,
right=28mm,
top=30mm,
bottom=30mm
]{geometry}

\usepackage[T1]{fontenc}
\usepackage[utf8]{inputenc}

\usepackage{lmodern}
\usepackage{microtype}

\usepackage{amsmath}
\usepackage{amsthm}
\usepackage{amsfonts}
\usepackage{amssymb}
\usepackage{mathtools}

\usepackage{mathrsfs}
\usepackage{bm}

\usepackage{tensor}
\usepackage{slashed}

\usepackage[svgnames]{xcolor}

\definecolor{LinkColor}{HTML}{1A4F8B}
\definecolor{CiteColor}{HTML}{2E6F40}
\definecolor{URLColor}{HTML}{7A3E9D}

\usepackage[
colorlinks=true,
linkcolor=LinkColor,
citecolor=CiteColor,
urlcolor=URLColor
]{hyperref}

\usepackage[nameinlink,capitalise]{cleveref}

\usepackage{graphicx}
\usepackage{subcaption}

\usepackage{tikz}
\usetikzlibrary{calc}

\usepackage{float}

\usepackage{diagbox}
\usepackage{makecell}

\numberwithin{equation}{section}

\theoremstyle{plain}

\theoremstyle{definition}

\theoremstyle{remark}

\newcommand{\mc}{\mathcal}
\newcommand{\mf}{\mathfrak}

\usepackage[
    backend=biber,
    style=numeric-comp,
    sorting=none,
    sortcites=true,
    maxnames=99,
    minnames=99,
    giveninits=true,
    doi=true,
    url=true,
    isbn=false,
    eprint=true
]{biblatex}

\renewbibmacro{in:}{}

\DeclareFieldFormat{doi}{}
\DeclareFieldFormat{url}{}

\DeclareFieldFormat[article]{journaltitle}{\mkbibemph{#1}}
\DeclareFieldFormat[article]{volume}{\mkbibbold{#1}}
\DeclareFieldFormat[article]{year}{\mkbibparens{#1}}
\AtEveryBibitem{%
    \clearfield{number}%
    \clearfield{issue}%
    \clearfield{month}%
    \clearfield{day}%
    \clearfield{date}%
}

\DeclareFieldFormat{pages}{#1}

\DeclareFieldFormat[article]
{title}{%
    \iffieldundef{doi}{%
        \iffieldundef{eprint}{%
            \iffieldundef{url}{%
                \mkbibemph{#1}%
            }{%
                \href{\thefield{url}}
                     {\mkbibemph{#1}}%
            }%
        }{%
            \href{https://arxiv.org/abs/\thefield{eprint}}
                 {\mkbibemph{#1}}%
        }%
    }{%
        \href{https://doi.org/\thefield{doi}}
             {\mkbibemph{#1}}%
    }%
}

\DeclareFieldFormat{eprint:arxiv}{%
    \mkbibbrackets{%
        \href{https://arxiv.org/abs/#1}
             {arXiv:#1}%
    }%
}

\title{\textbf{Connecting the tensor-categorical formulation of anyon condensation with operator algebras and entropic order parameters}}

\author{
Hua-Chen Zhang\\
\\
{\it Asia Pacific Center for Theoretical Physics,}\\
{\it 77 Cheongam-ro, Nam-gu, 37673 Pohang, Korea}\\
\\
\texttt{huachen.zhang@apctp.org}
}

\date{\today}

\begin{document}

\maketitle

\begin{abstract}
Anyon condensation that describes the transition between topological quantum field theories can be formulated in the language of tensor categories or that of operator algebras. We deploy the formalism of Doplicher-Haag-Roberts bimodules over quasi-local $\mathrm{C}^{*}$-algebras recently developed in~\cite{jones2024} to investigate anyon condensation, which is associated with an extension of a certain operator algebra. The connection between notions in the two formulations is thereby made manifest in an intuitive, diagrammatic manner. An entropic order parameter, as the quantum information-theoretic measure characterising a condensation, is naturally defined, and we give a very simple proof of a bound on it.
\end{abstract}

\tableofcontents

\newpage

\section{Introduction and summary}
\label{sec:intro}

Anyon condensation~\cite{bais2009,burnell2018,simon2023} is a mechanism that is believed to describe general transitions between (2+1)-dimensional topological orders~\cite{wen2017}, or equivalently, topological quantum field theories (TQFTs). Physically, as its name suggests, anyon condensation can be thought of as a conceptual extension of Bose-Einstein condensation, where (free) bosons are `absorbed' into the vacuum by tuning certain parameters of the Hamiltonian. The reason why the particles that can undergo condensation are bosons is that otherwise, an unphysical phase would be produced when the particles in the condensate move around each other. In anyon condensation, these are bosonic anyons in a topological order, i.e., those with trivial spin. An anyon type is then identified with the same type `dressed by' a condensate; sometimes, to fulfil consistency with the fusion rules, an irreducible anyon type in the original topological order must split into several ones in the post-condensation theory. Moreover, as the condensed anyons become invisible, it is clear that to define a theory with consistent braiding, other anyons that braid non-trivially with the condensed ones have to be removed from the system, or~\emph{confined}. These aspects were examined in depth in ref.~\cite{neupert2016}, where an algorithm for systematically constructing the anyon condensations of a given topological order was put forward.

The study of anyon condensation is a field where the beautiful interplay between physical principles and mathematical structures flourishes. It has now been well perceived that the data of a topological order, or TQFT, are captured by a (unitary) modular tensor category (MTC)~\cite{kitaev2006,turaev2016,turaev2017}. In light of this, one expects that anyon condensation is characterised by structures in an MTC. Indeed, anyon condensation has been formulated elegantly in the language of tensor categories~\cite{kong2014,eliens2014}, where a key conclusion is that~\emph{the condensed anyons form a connected commutative symmetric special Frobenius algebra object in the MTC describing the original topological order}; such an algebra object is termed a~\emph{condensable algebra}. The post-condensation theory is again described by an MTC, which is also determined in terms of the condensable algebra. On the other hand, the same structure emerges in the context of~\emph{subfactor theory}~\cite{jones1983,kosaki1986,longo1989,kosaki1992,longo1994,longo1995,popa1995,evans1998,izumi2000,mueger2003a,mueger2003b,bischoff2015,chen2022a,chen2022b,evans2023}, where the counterpart of a condensable algebra is known as a~\emph{commutative Q-system}. A factor here refers to a von Neumann algebra with trivial centre, and a Q-system is a certain Frobenius algebra object in the category of finite-dimensional endomorphisms over the factor. A Q-system characterises an extension of the factor, namely, an inclusion of the factor into a larger one determined by the Q-system. In the operator-algebraic formulation of quantum field theories~\cite{haag1964,haag1996}, the local observables generate (type III) factors, and equivalence classes of the endomorphisms are called Doplicher-Haag-Roberts (DHR) superselection sectors~\cite{doplicher1969a,doplicher1969b,doplicher1971,doplicher1974}; the latter are simple objects of an MTC under mild assumptions. In particular, commutative Q-systems correspond to local extensions of the operator algebra underlying the field theory. One advantage of the operator-algebraic formalism is that it emphasises the role of~\emph{quantum states}; consequently, quantum information-theoretic quantities are conveniently accessed.

Given this connection, one naturally anticipates that anyon condensation can also be formulated using the extension of operator algebras.\footnote{See, e.g., refs.~\cite{kawahigashi2021,vadnerkar2026} for reviews on the operator-algebraic approach to topological orders.} In symmetry-breaking scenarios, one usually has a pair of operator algebras where certain non-local operators are excluded or included, respectively. Consequently, an~\emph{entropic order parameter}~\cite{casini2020,magan2021,casini2021,magan2021a,molina2024,ahmad2026,zhang2026}, or entanglement asymmetry~\cite{marvian2014,capizzi2023,fossati2024,kusuki2025,fossati2025,ares2023,benini2025,lamas2026,benini2026,travaglino2026,vescovo2026}, can be identified as the relative entropy between a quantum state and its symmetrised counterpart. Very recently, anyon condensation was exploited from the perspective of generalised symmetry breaking~\cite{molina2026}, where the entropic order parameter was computed in the framework of finite-dimensional $\mathrm{C}^{*}$-algebras. A bound on this entropic order parameter was related to an~\emph{index} measuring the relative size of the pertinent algebras, which, in turn, coincides with the quantum dimension of the condensable algebra. More generally, given states on an operator algebra and its extension to a larger one, an entropic order parameter is defined as the relative entropy between them. In the present work, we elaborate on the mutually complementary formulations of anyon condensation based on operator algebras and tensor categories, thereby rendering the connection between notions in the two approaches manifest. Although various mathematical concepts are involved, our emphasis is not on rigorous proofs, and we hope to make our exposition more accessible to physicists through intuitive, diagrammatic arguments. In particular, the entropic order parameter for a condensation is naturally defined and computed, with the origin of its bound clearly exhibited.

The operator-algebraic formalism we deploy, which is different from that utilised in ref.~\cite{molina2026}, is based on the recently developed theory of DHR bimodules~\cite{jones2024}, where one considers the category $\mathrm{DHR}(\mf{M})$ of localisable bimodules over a quasi-local $\mathrm{C}^{*}$-algebra $\mf{M}$. Generalising the `traditional' DHR theory of endomorphisms, this approach is well suited to the situation where $\mf{M}$ is a so-called~\emph{fusion spin chain}, which is defined abstractly in terms of a fusion category and a specific object therein~\cite{hataishi2025}. The symmetry category of the algebra $\mf{M}$ is Morita equivalent to the fusion category from which $\mf{M}$ is defined. A key result of ref.~\cite{jones2024} is that $\mathrm{DHR}(\mf{M})$~\emph{is braided equivalent to the Drinfeld centre of this fusion category}; instead of repeating the proof, we will justify this claim by diagrammatic arguments. The most straightforward example is the representation category $\mathrm{Rep}(G)$ of a finite group $G$, where the regular representation is chosen as the specific object; the resulting quasi-local algebra is that of operators invariant under $G$, and the DHR category is simply the quantum double $\mathcal{Z}(\mathrm{Rep}(G))$. One immediately realises that $\mathrm{DHR}(\mf{M})$ describes the symmetry topological field theory (SymTFT)~\cite{ji2020,kong2020,gaiotto2021,burbano2022,chatterjee2023a,freed2024,kaidi2023a,kaidi2023b,bhardwaj2025} that encodes the categorical symmetry of $\mf{M}$~\cite{evans2026}.\footnote{Note that the Drinfeld centres of Morita equivalent fusion categories are braided equivalent~\cite{etingof2015}; in physical terms, one says that mutually dual categorical symmetries share the same SymTFT.} In fact, as we shall see below, the mathematical structures and diagrammatics used in our discussions parallel those in SymTFT; however, the logical perspectives are rather different: instead of obtaining the SymTFT from a given symmetric operator algebra, our starting point is the bulk topological order, and its associated operator algebra $\mf{M}$ is constructed as a fusion spin chain using the fusion category of defect lines on a topological boundary~\cite{jones2025,schatz2025,jones2026}. These aspects will be fleshed out in section~\ref{sec:formalism}; at this point, let us remark that by assuming the existence of a topological (i.e., `gapped') boundary, we restrict ourselves to the special case where the MTC describing the topological order is of trivial Witt class~\cite{davydov2013,fuchs2013}. Interesting examples including fractional quantum Hall states and chiral spin liquids are beyond this case\footnote{These~\emph{chiral topological states} admit robust gapless chiral edge modes described by chiral conformal field theory. The number of these edge modes is counted by the~\emph{chiral central charge}. That the latter vanishes ($\mathrm{mod}~8$) is a necessary but not sufficient condition for an MTC to be of trivial Witt class.}; a more general construction incorporating them will be considered in the future.

Using the language of operator algebras, a state $\omega$ on the $\mathrm{C}^*$-algebra $\mf{M}$ is a normalised positive linear functional $\omega: \mf{M} \rightarrow \mathbb{C}$, which manifests itself as a boundary of the TQFT described by the MTC $\mathrm{DHR}(\mf{M})$; indeed, given a reference topological boundary condition (e.g., the symmetry boundary of a SymTFT), an expectation value is assigned to the insertion of a local operator on the boundary labelled by $\omega$. In particular, a state corresponding to a topological/gapped boundary is also termed topological. As will be elaborated on below, a condensable algebra object $A$ in $\mathrm{DHR}(\mf{M})$ defines an irreducible local inclusion of $\mf{M}$ into a larger algebra $\mf{A}$, and the MTC for the `post-condensation' theory is simply $\mathrm{DHR}(\mf{A})$. The inclusion is associated with a unique~\emph{conditional expectation} $E: \mf{A} \rightarrow \mf{M}$, which is a positive linear unital map satisfying the bimodule property
\begin{equation}
\label{eq:bimodule-property}
    E(\mc{O}_{1} \widetilde{\mc{O}} \mc{O}_{2}) = \mc{O}_{1} E(\widetilde{\mc{O}}) \mc{O}_{2}, \quad \forall~\mc{O}_{1}, \mc{O}_{2} \in \mf{M}~\text{and}~\widetilde{\mc{O}} \in \mf{A}.
\end{equation}
Composing $\omega$ with the conditional expectation, one obtains a state $\omega \circ E$ on $\mf{A}$; this will have a clear meaning in our diagrammatics. For each pure extension $\omega_{r}$ of a topological state $\omega$ on $\mathfrak{A}$, the value of the entropic order parameter is defined as the relative entropy
\begin{equation}
    S(\omega_{r} | \omega \circ E) = \mathrm{Tr}\left[ \rho_{\omega_{r}} \left( \log{\rho_{\omega_{r}}} - \log{\rho_{\omega \circ E}} \right) \right]
\end{equation}
with $\rho_{\omega}$ the density matrix of $\omega$. Here, both $\rho_{\omega_{r}}$ and $\rho_{\omega \circ E}$ are represented on the space of genuinely local operators in $\mf{A}$ (see below), which is finite-dimensional. For a pure state $\omega$ on $\mf{M}$ and any extension $\omega_{r}$ on $\mf{A}$, there is a bound for this entropic order parameter given by the quantum dimension $d_{A}$ of the condensable algebra $A$:
\begin{equation}
    S(\omega_{r} | \omega \circ E) \leq \log{d_{A}}.
\end{equation}
This result agrees with the Pimsner-Popa bound~\cite{pimsner1986}, which has been extended to the context of $\mathrm{C}^*$-algebras, where $d_{A}$ is identified as the Watatani index~\cite{watatani1990} of the corresponding inclusion. By combining the operator-algebraic and the tensor-categorical approaches, we will give a very simple proof of this bound.

The remaining part of the text begins by reviewing the tensor-categorical approach to anyon condensation (subsection~\ref{subsec:review}). The diagrammatics connecting the tensor-categorical and the operator-algebraic formulations is developed mainly in subsection~\ref{subsec:connecting}. In subsection~\ref{subsec:entropic-order-parameter}, the states appearing in the entropic order parameter for a condensation are defined, and the bound is proven. In section~\ref{sec:examples}, the entropic order parameters for several simple examples of anyon condensation are computed. The mathematical definition and some properties of a condensable algebra are collected in appendix~\ref{sec:condensable-algebras} for completeness.

\section{Formalism}
\label{sec:formalism}

\subsection{Review of tensor-categorical approach to anyon condensation}
\label{subsec:review}

We begin with a minimal review of the tensor-categorical approach to anyon condensation, while setting up our notations along the way. For a detailed account, we refer the reader to ref.~\cite{kong2014}.

\begin{figure}[htbp]
\centering
\includegraphics[width=\textwidth]{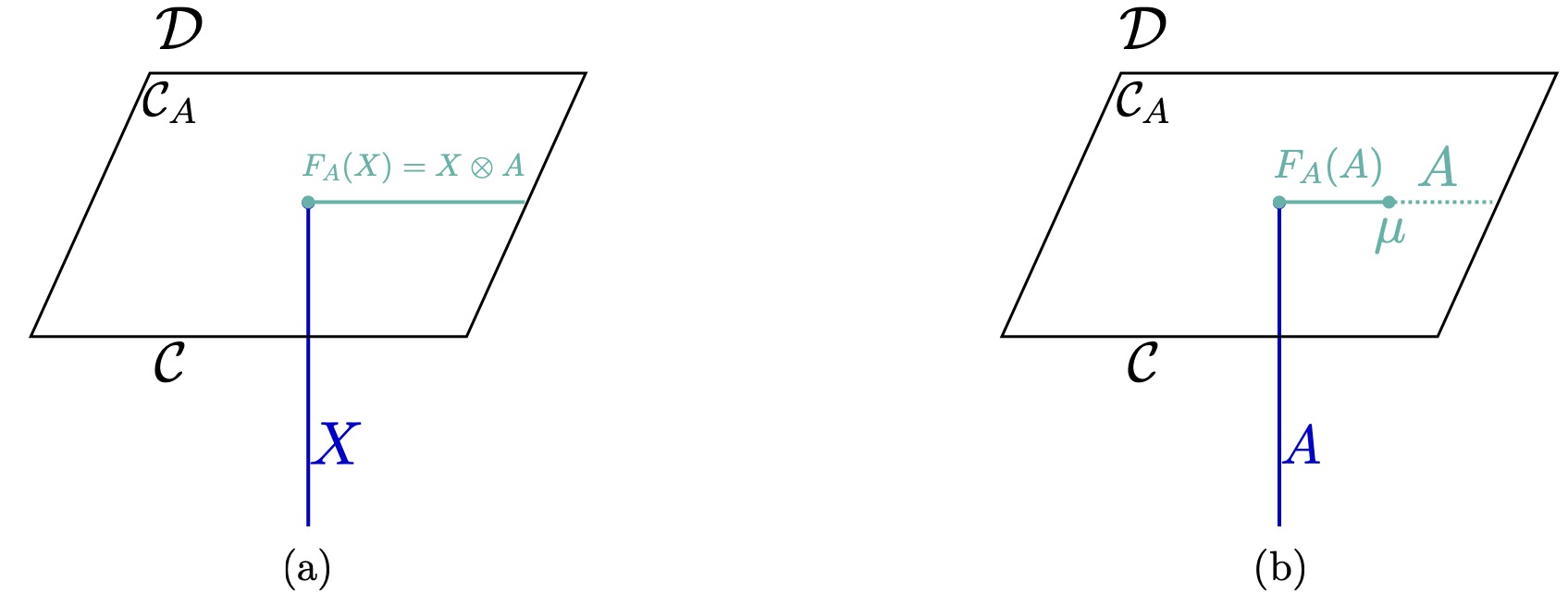}
\caption{(a) The free $A$-module functor $F_{A}:~\mathcal{C} \rightarrow \mathcal{C}_{A}$ with $F_{A}(X) = X \otimes A$ determines what happens when an anyon $X$ from topological order $\mathcal{C}$ is brought to the interface described by the module category $\mathcal{C}_{A}$. $\mathcal{D}$, sitting on the other side of this interface, is the post-condensation topological order to be identified below. (b) As $A$ is the invisible topological line (depicted as a dashed line here and simply omitted in the following) on the interface $\mathcal{C}_{A}$, composing $F_{A}$ with the multiplication $\mu: A \otimes A \rightarrow A$, one finds that the~\emph{bulk} line $A$ can end topologically on this interface.~\label{fig:1}}
\end{figure}

Suppose we are given a (2+1)-dimensional topological order, or TQFT, described by an MTC $\mathcal{C}$; the simple objects of $\mathcal{C}$ are in one-to-one correspondence with the irreducible anyon types, and label the simple topological lines in this TQFT. A condensable algebra $A$ is a (not necessarily simple) connected commutative symmetric special Frobenius algebra object, for which the mathematical definition is provided in appendix~\ref{sec:condensable-algebras}. `Making the condensable algebra invisible' amounts to finding objects that can `absorb' the action of $A$; mathematically, these are modelled by the $A$-\emph{modules} in $\mathcal{C}$. Without loss of generality, we assume that $A$ acts from the right side, i.e., there is a morphism $m \otimes A \rightarrow m$ satisfying certain consistency conditions which we omit here~\cite{etingof2015}. Such objects $m$ equipped with the right $A$ action form the category of right $A$-modules, denoted as $\mathcal{C}_{A}$. The latter is an indecomposable module category over $\mathcal{C}$. Physically, the module category $\mathcal{C}_{A}$ describes a topological (or `gapped') interface in the (2+1)d spacetime, where the TQFT $\mathcal{C}$ lives in the half on one side of this interface. On the latter, the topological lines are labelled by the objects in $\mathcal{C}_{A}$. To describe what happens when a bulk topological line (i.e., an anyon) $X$ in $\mathcal{C}$ is brought to the interface, as shown in figure~\ref{fig:1}(a)\footnote{A subtlety is that for an anyon which is not self-dual, one generally needs to specify the orientation of the corresponding topological line by putting an arrow on it. As it will not cause confusion, we ignore this subtlety here and throughout the text.}, we have a central functor, the free $A$-module functor $F_{A}:~\mathcal{C} \rightarrow \mathcal{C}_{A}$ with $F_{A}(X) = X \otimes A$. Note that $X \otimes A$ is a right $A$-module thanks to the structure of multiplication $\mu: A \otimes A \rightarrow A$; particularly, $A$ is itself a right $A$-module. In fact, any object $m$ in $\mathcal{C}_{A}$ is also an $(A,A)$-\emph{bimodule}, i.e., admits consistent $A$ actions from both sides. This is because a left $A$ action on $m$ can be induced from the right $A$ action using the braiding in $\mathcal{C}$, a process known as $\alpha$-induction~\cite{longo1995} from the study of subfactors; the consistency between the left and right actions is guaranteed by the commutativity of $A$. Therefore, one can define a fusion on $\mathcal{C}_{A}$ as the tensor product over $A$, $\otimes_{A}$~\cite{etingof2015}. It is easily proved that $A \otimes_{A} m \cong m \otimes_{A} A \cong m$ for any $m \in \mathcal{C}_{A}$. This means the topological line $A$ on the interface is the tensor unit with respect to $\otimes_{A}$, in agreement with our physical intuition that $A$ becomes invisible after condensation. Moreover, one verifies immediately that the bulk line $A$ in $\mathcal{C}$ can end topologically on this interface: $F_{A}(A) = A \otimes A \xrightarrow{\mu} A$; see figure~\ref{fig:1}(b).

Crucially, however, the left $A$ action on $m$ can be defined in two ways by $\alpha$-induction, using either the `over braiding' $c_{A,m}$ or the `under braiding' $c_{m,A}^{-1}$. Generally, these result in two different bimodule structures on $m$, implying that the corresponding particle braids non-trivially with the condensate $A$. As we argued in section~\ref{sec:intro}, such particles must be confined. More concretely, they are confined on the interface described by $\mathcal{C}_{A}$; indeed, for topological lines living on the interface, despite they can fuse with each other, a braiding is not defined. The deconfined particles, on the other hand, correspond to the modules for which the two left actions coincide. These are termed the~\emph{local} $A$-modules; the latter form a full subcategory of $\mathcal{C}_{A}$ denoted as $\mathcal{C}_{A}^{\mathrm{loc}}$, on which a consistent braiding can be defined, making it an MTC. The deconfined particles are the only ones that can leave the interface and become bulk anyons in the post-condensation topological order $\mathcal{D}$ occupying the spacetime on the other side of this interface. Namely, we have $\mathcal{D} = \mathcal{C}_{A}^{\mathrm{loc}}$. We are thus led to the following punchline:\footnote{In the seminal paper~\cite{bais2009} where anyon condensation was put forward, the original theory $\mathcal{C}$, the theory $\mathcal{C}_{A}$ with deconfined~\emph{and} confined particles, and the post-condensation theory $\mathcal{C}_{A}^{\mathrm{loc}}$ were denoted as $\mathcal{A}$, $\mathcal{T}$, and $\mathcal{U}$, respectively. We emphasise again that whereas $\mathcal{A}$ and $\mathcal{U}$ are topological orders (MTCs), $\mathcal{T}$ is only a fusion theory without a consistent braiding.}

\begin{equation*}
\begin{tikzpicture}[
    scale=1.15,
    line join=round,
    line cap=round
]

\def\L{5.4}
\def\H{1.2}
\def\D{0.65}
\def\M{2.7}

\filldraw[
fill=MediumBlue,
fill opacity=0.72,
draw=black,
line width=0.45pt]
(0,0) rectangle (\M,\H);

\filldraw[
fill=DarkSeaGreen,
fill opacity=0.72,
draw=black,
line width=0.45pt]
(\M,0) rectangle (\L,\H);

\filldraw[
fill=MediumBlue,
fill opacity=0.72,
draw=black,
line width=0.45pt]
(0,\H) --
(\M,\H) --
(\M+\D,\H+\D) --
(\D,\H+\D) --
cycle;

\filldraw[
fill=DarkSeaGreen,
fill opacity=0.62,
draw=black,
line width=0.45pt]
(\M,\H) --
(\L,\H) --
(\L+\D,\H+\D) --
(\M+\D,\H+\D) --
cycle;

\filldraw[
fill=DarkSeaGreen,
fill opacity=0.62,
draw=black,
line width=0.45pt]
(\L,0) --
(\L,\H) --
(\L+\D,\H+\D) --
(\L+\D,\D) --
cycle;

\filldraw[
fill=LightSeaGreen,
fill opacity=0.62,
draw=black,
line width=0.45pt]
(\M,0) --
(\M,\H) --
(\M+\D,\H+\D) --
(\M+\D,\D) --
cycle;

\draw[densely dashed,gray]
(\D,\D)--(\M+\D,\D);

\draw[densely dashed,gray]
(\M+\D,\D)--(\L+\D,\D);

\draw[densely dashed,gray]
(\D,\D)--(\D,\H+\D);

\draw[line width=0.45pt] (0,0)--(\L,0);
\draw[line width=0.45pt] (0,\H)--(\L,\H);
\draw[line width=0.45pt] (0,0)--(0,\H);
\draw[line width=0.45pt] (\L,0)--(\L,\H);

\draw[densely dashed,gray]
(0,0)--(\D,\D);

\draw[line width=0.45pt]
(\D,\H+\D)--(\L+\D,\H+\D);

\draw[line width=0.45pt]
(\L,\H)--(\L+\D,\H+\D);

\draw[line width=0.45pt]
(\L+\D,\H+\D)--(\L+\D,\D);

\draw[line width=0.45pt]
(\L+\D,\D)--(\L,0);

\draw[line width=0.45pt]
(0,\H)--(\D,\H+\D);

\draw[line width=0.45pt]
(\M,\H)--(\M+\D,\H+\D);

\draw[densely dashed,gray]
(\M,0)--(\M+\D,\D);

\draw[densely dashed,gray]
(\M+\D,\D)--(\M+\D,\D+\H);

\node at (1.35,0.90)
{$\displaystyle \mathcal{C}$};

\node at (3.00,0.90)
{\small$\displaystyle \mathcal{C}_{A}$};

\node at (4.20,0.90)
{$\displaystyle \mathcal{C}_{A}^{\mathrm{loc}}$};

\end{tikzpicture}
\end{equation*}

\emph{For a topological order described by an MTC $\mathcal{C}$, an anyon condensation is defined in terms of a condensable algebra $A \in \mathcal{C}$.\footnote{Of course, condensable algebras that are Morita equivalent lead to equivalent anyon condensations.} The post-condensation topological order consists of the~\emph{deconfined} particles, which correspond to the~\emph{local} A-modules forming the MTC $\mathcal{C}_{A}^{\mathrm{loc}}$. The category of $A$-modules, $\mathcal{C}_{A}$, describes the topological interface between the original and post-condensation theories, on which both confined and deconfined particles can live.}

The total quantum dimension of an MTC $\mc{C}$ is defined in terms of the quantum dimension $d_{a}$ of each simple object $a \in \mathrm{Irr}(\mc{C})$ [$\mathrm{Irr}(\mc{C})$ denotes the set of simple objects in $\mc{C}$] as $\mathrm{dim}(\mathcal{C}) = \sum_{a \in \mathrm{Irr}(\mc{C})} d_{a}^{2}$. It can be shown~\cite{froehlich2006,simon2023} that the total quantum dimensions of the original and post-condensation topological orders are related via
\begin{equation}
\label{eq:relation-quantum-dimensions}
    \frac{\mathrm{dim}(\mathcal{C})}{\mathrm{dim}(\mathcal{C}_{A}^{\mathrm{loc}})} = d_{A}^{2}.
\end{equation}
As an important example, a~\emph{Lagrangian algebra $L$} is a condensable algebra object in $\mathcal{C}$ with quantum dimension $d_{L} = \mathrm{dim}(\mathcal{C})^{1/2}$. Relation~\eqref{eq:relation-quantum-dimensions} then implies that $\mathcal{C}_{L}^{\mathrm{loc}} \cong \mathrm{Vec}$ (the category of finite-dimensional vector spaces). This means that the anyon condensation associated with a Lagrangian algebra is~\emph{fully confining}, namely, the post-condensation topological order is trivial. In this case, $\mathcal{C}_{L}$ describes a topological~\emph{boundary} of the TQFT $\mathcal{C}$.

\subsection{Connecting with operator algebras}
\label{subsec:connecting}

Having reviewed ingredients in the description of anyon condensation with the tensor-categorical language, our aim in this subsection is to establish a complementary formalism based on operator algebras. Instead of presenting rigorous proofs, we will employ intuitive diagrammatic arguments to make the mathematical concepts less abstract; comparing with the results of subsection~\ref{subsec:review} immediately renders the correspondence between notions in the two approaches clear.

In the operator-algebraic approach to quantum field theories~\cite{haag1996}, a theory is defined by assigning an algebra of local observables to each open spacetime region, resulting in a~\emph{causal net} of operator algebras, for which the representations are encoded in the so-called DHR endomorphisms. Recently, the DHR theory was adapted to describe abstract spin chains, where the endomorphisms are replaced by~\emph{bimodules} over the pertinent $\mathrm{C}^{*}$-algebra~\cite{jones2024}. In the following, we apply this formalism to the situation of anyon condensation.

To identify the net of $\mathrm{C}^{*}$-algebras associated with the topological order described by $\mathcal{C}$, we pick an arbitrary Lagrangian algebra $L \in \mathcal{C}$. As we noted at the end of subsection~\ref{subsec:review}, $\mathcal{C}_{L}$ describes a topological boundary of $\mathcal{C}$, on which the topological lines are labelled by the objects in $\mathcal{C}_{L}$. Suppose $M \in \mathcal{C}_{L}$ is a self-dual object. It is~\emph{strongly tensor generating} if each (isomorphism class of) simple object in $\mathcal{C}_{L}$ appears in the decomposition of $M^{\otimes l}$ for the positive integral power $l$ large enough. The idea is to interpret $M$ as an `on-site Hilbert space', and the algebra of operators acting on it is hence given by the morphism space $\mathrm{End}_{\mathcal{C}_{L}}(M) \equiv \mathrm{Hom}_{\mathcal{C}_{L}}(M,M)$, which admits the structure of a $\mathrm{C}^{*}$-algebra. Therefore, assigning $\mf{M}_{l} \equiv \mathrm{End}_{\mathcal{C}_{L}}(M^{\otimes l})$ to a segment containing $l$ `sites' yields a discrete net of $\mathrm{C}^{*}$-algebras\footnote{Such a net satisfies natural requirements, e.g. the algebraic Haag duality~\cite{jones2024}.}:
\begin{equation*} \includegraphics[width=0.5\textwidth]{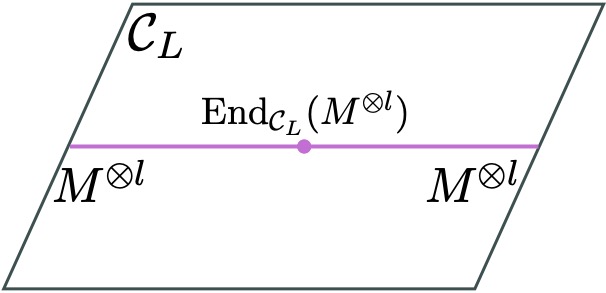}
\end{equation*}
Note that in the visualisation of the morphism space, $l$ sites have been blocked together. In fact, the quasi-local $\mathrm{C}^{*}$-algebra $\mf{M} := \lim\limits_{\to} \mf{M}_{l}$, known as a fusion spin chain, is well defined as the inductive limit of this sequence. We call $\mf{M}$ a~\emph{boundary algebra} of the TQFT $\mc{C}$.

\begin{figure}[htbp]
\centering
\includegraphics[width=\textwidth]{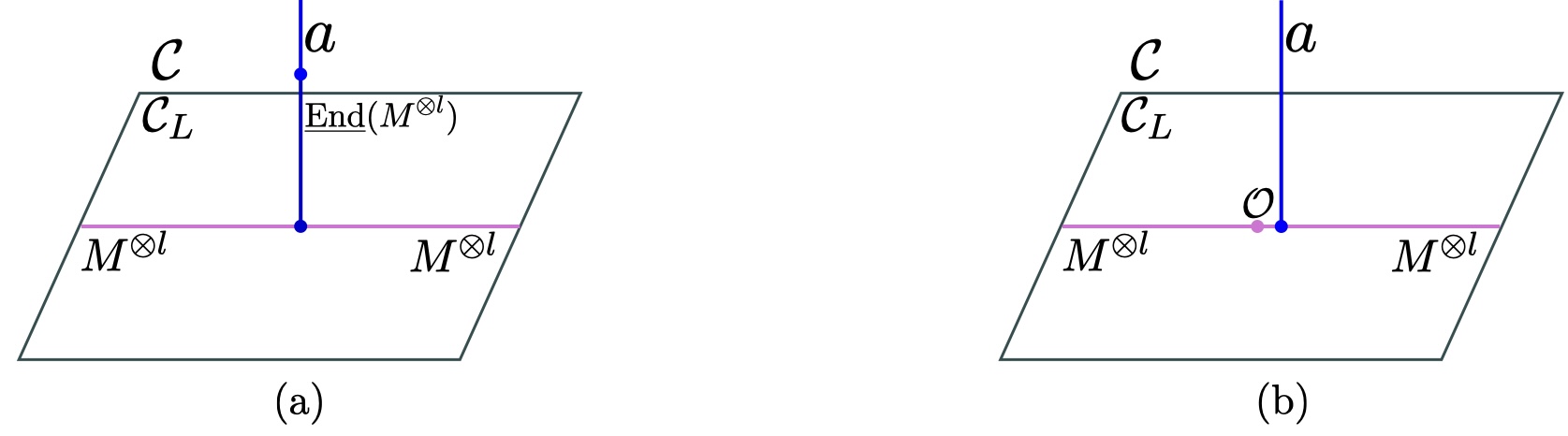}
\caption{(a) For a strongly tensor generating object $M \in \mc{C}_{L}$ and a sufficiently large $l$, any $a \in \mathrm{Irr}(\mc{C})$ arises in the decomposition of the internal end $\underline{\mathrm{End}}(M^{\otimes l})$. (b) Acting an operator $\mathcal{O} \in \mathrm{End}_{\mathcal{C}_{L}}(M^{\otimes l})$ on an element in the bimodule labelled by $a$, which is represented as a junction operator joining the bulk line $a$ with the boundary lines $M^{\otimes l}$, amounts to bringing $\mathcal{O}$ close to the junction and performing the OPE. The latter produces another junction operator attached to $a$.~\label{fig:2}}
\end{figure}

Now, we would like to come up with a diagrammatic representation of the objects in $\mathrm{DHR}(\mf{M})$, which are localisable\footnote{Roughly, localisability means that the elements are invariant under the action of local operators outside of a certain finite region; clearly, this region is given by the segment of the fusion spin chain with length $l$ in our setup.} bimodules over $\mf{M}$. To this end, we use the notion of the~\emph{internal hom}~\cite{etingof2015} $\underline{\mathrm{Hom}}(M,N)$ between objects $M$ and $N$ in $\mc{C}_{L}$. As an object in $\mc{C}$, it is defined by the isomorphism
\begin{equation}
\label{eq:internal-hom-def}
    \mathrm{Hom}_{\mc{C}}(X, \underline{\mathrm{Hom}}(M,N)) \cong \mathrm{Hom}_{\mc{C}_{L}}(X \otimes M, N)
\end{equation}
$\forall X \in \mc{C}$, where the $\otimes$ on the right hand side is the action of $\mc{C}$ on $\mc{C}_{L}$, which is a module category over $\mc{C}$. The internal hom is represented as a bulk topological line:
\begin{equation*}
\includegraphics[width=0.5\textwidth]{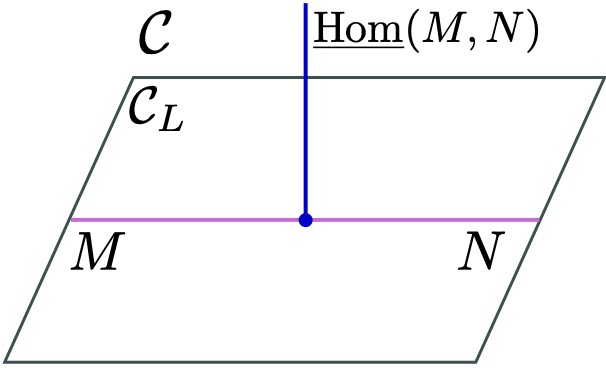}
\end{equation*}
Using~\eqref{eq:internal-hom-def}, one finds that
\begin{equation} \mathrm{dim}\mathrm{Hom}_{\mc{C}}(a, \underline{\mathrm{End}}(M^{\otimes l})) = \mathrm{dim}\mathrm{Hom}_{\mc{C}_{L}}(a \otimes M^{\otimes l}, M^{\otimes l}) \quad \forall a \in \mathrm{Irr}(\mc{C}).
\end{equation}
If $M$ is strongly tensor generating, by definition, the right hand side of the above equation is non-zero for arbitrary positive integer $l$ that is sufficiently large, implying that any simple object $a \in \mathrm{Irr}(\mc{C})$ can arise in the decomposition of the internal end; see figure~\ref{fig:2}(a). We claim that any bulk topological line $a$ labels a bimodule over the algebra $\mf{M}$, and the elements in this bimodule are simply the junction operators joining the bulk line $a$ with the boundary lines $M^{\otimes l}$. Indeed, the action of an operator $\mathcal{O} \in \mf{M}_{l} = \mathrm{End}_{\mathcal{C}_{L}}(M^{\otimes l})$ amounts to bringing $\mathcal{O}$ close to the junction and performing the operator product expansion (OPE) with the junction operator, which is depicted in figure~\ref{fig:2}(b). Clearly, this OPE leaves the bulk line intact, and the outcome is another junction operator attached again to $a$. The operator can act from the other side of the junction as well, endowing $a$ with the structure of a bimodule over $\mf{M}_{l}$. Finally, taking the inductive limit, this argument carries over to the quasi-local $\mathrm{C}^{*}$-algebra $\mf{M}$. It is also intuitively clear that any localisable bimodule over $\mf{M}$ can be obtained in this way. We have thus argued in an intuitive, diagrammatic manner that $\mc{C} \cong \mathrm{DHR}(\mf{M})$, a key result proved in ref.~\cite{jones2024}; this construction does not depend on the choice of the Lagrangian algebra $L$ or the strongly tensor generating object $M$.

\begin{figure}[htbp]
\centering
\includegraphics[width=\textwidth]{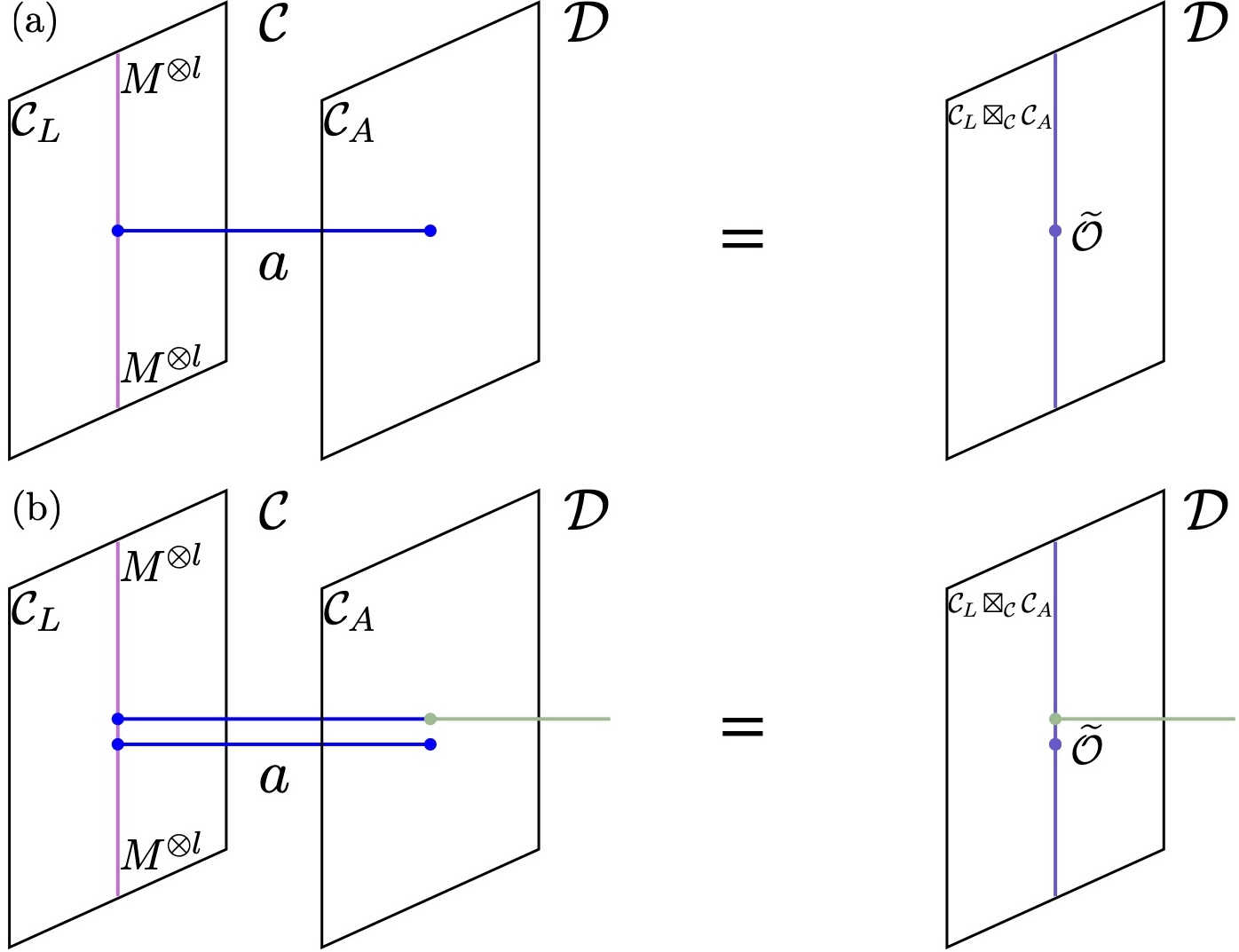}
\caption{The bulk line $a \in \mathrm{Irr}(\mc{C})$ can end topologically on the interface $\mc{C}_{A}$, and $M$ is a strongly tensor generating object in $\mc{C}_{L}$. (a) An operator sits at the junction joining $a$ with the lines $M^{\otimes l}$ on the boundary. $\widetilde{\mc{O}}$, obtained when hitting this junction operator with the interface by collapsing the $\mc{C}$ layer, becomes an element in $\mf{A}$ upon taking the inductive limit. (b) The action of $\widetilde{\mc{O}}$ via OPE. Repeating the arguments we made above [c.f. figure~\ref{fig:2}(b)], one realises that every bulk line in the TQFT $\mc{D}$ labels a bimodule over $\mf{A}$; in fact, $\mc{D} \cong \mathrm{DHR}(\mf{A})$.~\label{fig:3}}
\end{figure}

After the operator algebra associated with a topological order is identified, let us proceed to formulate anyon condensation in this framework. More specifically, we would like to determine the boundary algebra $\mf{A}$ for the post-condensation theory $\mc{D}$ corresponding to a condensable algebra $A$. This is again neatly accomplished by using the diagrammatic arguments.
\begin{equation*}
\raisebox{-0.5\height}{\begin{tikzpicture}[
    scale=1.15,
    line join=round,
    line cap=round
]

\def\L{5.4}
\def\H{1.2}
\def\D{0.65}
\def\M{2.7}

\filldraw[
fill=MediumBlue,
fill opacity=0.72,
draw=black,
line width=0.45pt]
(0,0) rectangle (\M,\H);

\filldraw[
fill=DarkSeaGreen,
fill opacity=0.72,
draw=black,
line width=0.45pt]
(\M,0) rectangle (\L,\H);

\filldraw[
fill=MediumBlue,
fill opacity=0.72,
draw=black,
line width=0.45pt]
(0,\H) --
(\M,\H) --
(\M+\D,\H+\D) --
(\D,\H+\D) --
cycle;

\filldraw[
fill=DarkSeaGreen,
fill opacity=0.62,
draw=black,
line width=0.45pt]
(\M,\H) --
(\L,\H) --
(\L+\D,\H+\D) --
(\M+\D,\H+\D) --
cycle;

\filldraw[
fill=DarkSeaGreen,
fill opacity=0.62,
draw=black,
line width=0.45pt]
(\L,0) --
(\L,\H) --
(\L+\D,\H+\D) --
(\L+\D,\D) --
cycle;

\filldraw[
fill=Orchid,
fill opacity=0.62,
draw=black,
line width=0.45pt]
(0,0) --
(0,\H) --
(0+\D,\H+\D) --
(0+\D,\D) --
cycle;

\filldraw[
fill=LightSeaGreen,
fill opacity=0.62,
draw=black,
line width=0.45pt]
(\M,0) --
(\M,\H) --
(\M+\D,\H+\D) --
(\M+\D,\D) --
cycle;

\draw[densely dashed,gray]
(\D,\D)--(\M+\D,\D);

\draw[densely dashed,gray]
(\M+\D,\D)--(\L+\D,\D);

\draw[densely dashed,gray]
(\D,\D)--(\D,\H+\D);

\draw[line width=0.45pt] (0,0)--(\L,0);
\draw[line width=0.45pt] (0,\H)--(\L,\H);
\draw[line width=0.45pt] (0,0)--(0,\H);
\draw[line width=0.45pt] (\L,0)--(\L,\H);

\draw[densely dashed,gray]
(0,0)--(\D,\D);

\draw[line width=0.45pt]
(\D,\H+\D)--(\L+\D,\H+\D);

\draw[line width=0.45pt]
(\L,\H)--(\L+\D,\H+\D);

\draw[line width=0.45pt]
(\L+\D,\H+\D)--(\L+\D,\D);

\draw[line width=0.45pt]
(\L+\D,\D)--(\L,0);

\draw[line width=0.45pt]
(0,\H)--(\D,\H+\D);

\draw[line width=0.45pt]
(\M,\H)--(\M+\D,\H+\D);

\draw[densely dashed,gray]
(\M,0)--(\M+\D,\D);

\draw[densely dashed,gray]
(\M+\D,\D)--(\M+\D,\D+\H);

\node at (0.25,0.90)
{\small$\displaystyle \mathcal{C}_{L}$};

\node at (1.35,0.90)
{$\displaystyle \mathcal{C}$};

\node at (3.00,0.90)
{\small$\displaystyle \mathcal{C}_{A}$};

\node at (4.20,0.90)
{$\displaystyle \mathcal{D}$};

\end{tikzpicture}
}
\quad \Rightarrow \quad
\raisebox{-0.5\height}{\begin{tikzpicture}[
    scale=1.15,
    line join=round,
    line cap=round
]

\def\L{5.4}
\def\H{1.2}
\def\D{0.65}
\def\M{2.7}

\filldraw[
fill=DarkSeaGreen,
fill opacity=0.62,
draw=black,
line width=0.45pt]
(0,0) rectangle (\M,\H);

\filldraw[
fill=DarkSeaGreen,
fill opacity=0.62,
draw=black,
line width=0.45pt]
(0,\H) --
(\M,\H) --
(\M+\D,\H+\D) --
(\D,\H+\D) --
cycle;

\filldraw[
fill=DarkSeaGreen,
fill opacity=0.62,
draw=black,
line width=0.45pt]
(\M,0) --
(\M,\H) --
(\M+\D,\H+\D) --
(\M+\D,\D) --
cycle;

\filldraw[
fill=SlateBlue,
fill opacity=0.62,
draw=black, densely dashed]
(0,0) --
(0,\H) --
(0+\D,\H+\D) --
(0+\D,\D) --
cycle;

\draw[densely dashed,gray]
(0,0)--(\D,\D);

\draw[densely dashed,gray]
(\D,\D)--(\M+\D,\D);

\draw[densely dashed,gray]
(\D,\D)--(\D,\H+\D);

\node at (0.05,0.90)
{\small$\mc{C}_{L} \boxtimes_{\mc{C}} \mc{C}_{A}$};

\node at (1.35,0.90)
{$\displaystyle \mathcal{D}$};

\end{tikzpicture}
}
\end{equation*}
As illustrated above\footnote{A configuration of TQFTs, boundaries and interfaces like this was called a `club quiche' in the context of SymTFT in refs.~\cite{bhardwaj2025c,bhardwaj2025d,bhardwaj2026}.}, a topological boundary of the TQFT $\mc{D}$ is produced by `collapsing' the theory $\mc{C}$ and hitting the boundary $\mc{C}_{L}$ with the interface $\mc{C}_{A}$. We denote the boundary thus obtained formally by $\mc{C}_{L} \boxtimes_{\mc{C}} \mc{C}_{A}$, which has the structure of a multi-fusion category. An operator in the algebra $\mf{A}$ is represented diagrammatically as in figure~\ref{fig:3}(a), where the simple bulk line $a \in \mathrm{Irr}(\mc{C})$ can end topologically on the interface $\mc{C}_{A}$. An operator sits at the junction joining $a$ with the lines $M^{\otimes l}$ on the boundary $\mc{C}_{L}$ ($M$ is strongly tensor generating as above).\footnote{A subtlety here is that we have implicitly assumed that a strongly tensor generating object in $\mc{C}_{L} \boxtimes_{\mc{C}} \mc{C}_{A}$ descends from $M$ in $\mc{C}_{L}$. We believe this to be the case, intuitively because the Lagrangian algebra $L$ is a `maximal' one among all the condensable algebras in $\mc{C}$.} The operator $\widetilde{\mc{O}}$ is obtained when hitting this junction operator with the interface by collapsing the $\mc{C}$ layer. $\widetilde{\mc{O}}$ becomes an element in $\mf{A}$ upon taking the inductive limit. It is readily seen that $\mf{M}$ is a subalgebra of $\mf{A}$, as the operators in $\mf{M}$ are nothing but those junction operators for which $a$ is the trivial bulk line. Repeating the argument we made for $\mf{M}$ [c.f. figure~\ref{fig:2}(b)], one realises that every bulk line in the topological order $\mc{D}$ labels a bimodule over $\mf{A}$; comparing with the results obtained in subsection~\ref{subsec:review} implies that $\mc{D} = \mc{C}_{A}^{\mathrm{loc}} \cong \mathrm{DHR}(\mf{A})$. The action of $\widetilde{\mc{O}}$ on the bimodule via OPE is illustrated in figure~\ref{fig:3}(b).\footnote{Notice that the diagrams in~figure~\ref{fig:3} are rotated by $90^{\circ}$ comparing to those in figure~\ref{fig:2} to save layout space.}

How to quantify the relative size of the extended algebra $\mf{A}$ comparing to $\mf{M}$? Our discussions above made it evident that this amounts to asking which bulk lines in $\mc{C}$ can end topologically on the interface $\mc{C}_{A}$; as we have seen in subsection~\ref{subsec:review}, they form the condensable algebra $A$. Concretely, if a simple line $a \in \mathrm{Irr}(\mc{C})$ can end on $\mc{C}_{A}$ in $n^{(A)}_{a}$ independent ways [namely, the space of junction operators is $n^{(A)}_{a}$-dimensional; $n^{(A)}_{a} = 0$ if $a$ cannot end topologically on $\mc{C}_{A}$], we express the condensable algebra as
\begin{equation}
\label{eq:condensable-algebra}
    A = \bigoplus_{a \in \mathrm{Irr}(\mc{C})} n^{(A)}_{a} a.
\end{equation}
As has been noted above, there is a unique conditional expectation $E: \mf{A} \rightarrow \mf{M}$ associated with the inclusion. We will see shortly that this allows for a natural definition of an entropic order parameter for the anyon condensation, which is directly related to the condensable algebra $A$; this is the goal of the next subsection.

\subsection{Entropic order parameter and its bound}
\label{subsec:entropic-order-parameter}

A state on the quasi-local $\mathrm{C}^{*}$-algebra $\mf{M}$ manifests itself as a boundary of the corresponding TQFT described by $\mc{C} \cong \mathrm{DHR}(\mf{M})$. In the following, we shall focus on~\emph{topological} states that correspond to topological boundaries; thus, a topological state $\omega$ is associated with a Lagrangian algebra $L_{\omega} \in \mc{C}$. To justify this, one recalls that the state is nothing but a normalised positive linear functional $\omega: \mf{M} \rightarrow \mathbb{C}$. By definition, the identity (i.e., the trivial line on the boundary) $\boldsymbol{1}$ appears in the decomposition of $M^{\otimes l}$ for $M \in \mc{C}_{L_{\omega}}$ a strongly tensor generating object and arbitrary positive integer $l$ that is large enough:
\begin{equation*}   \includegraphics[width=0.8\textwidth]{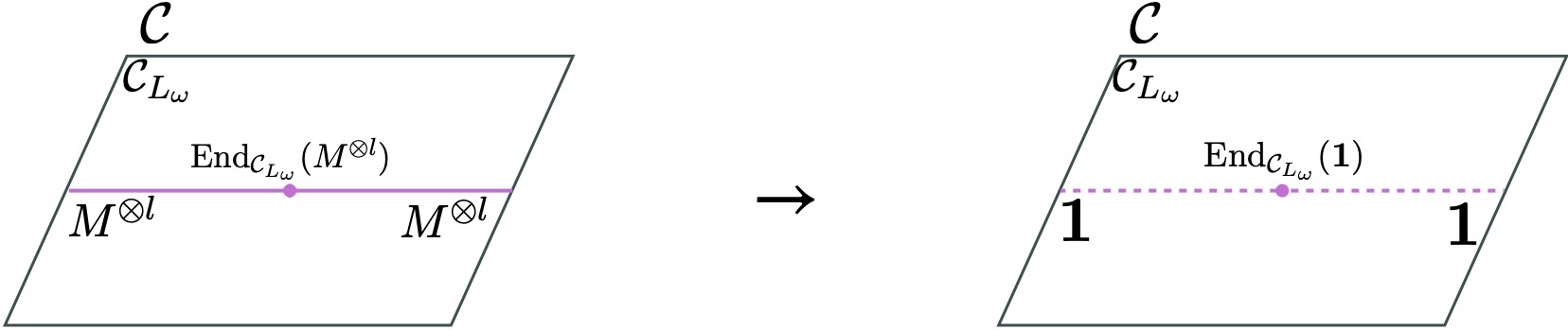}
\end{equation*}
Taking the inductive limit, an element in $\mf{M}$ produces one in $\mathrm{End}_{\mc{C}_{L_{\omega}}}(\boldsymbol{1})$, i.e., a~\emph{genuinely local} operator on the boundary $\mc{C}_{L_{\omega}}$, under this restriction. With an arbitrary reference topological boundary condition, the expectation value of this genuinely local operator is defined as the image of the linear functional $\omega$, justifying our claim. In particular, $\omega$ is a~\emph{pure} state as the topological boundary $\mc{C}_{L_{\omega}}$ is~\emph{irreducible}, namely $\mathrm{dim}\mathrm{End}_{\mc{C}_{L_{\omega}}}(\boldsymbol{1}) = 1$, since $L_{\omega}$ is~\emph{connected}; see appendix~\ref{sec:condensable-algebras}.

The conditional expectation $E: \mf{A} \rightarrow \mf{M}$, which lifts a state $\omega$ on $\mf{M}$ to $\omega \circ E$ on $\mf{A}$, is defined straightforwardly by the following diagrammatic equality:
\begin{equation*}
\includegraphics[width=0.8\textwidth]{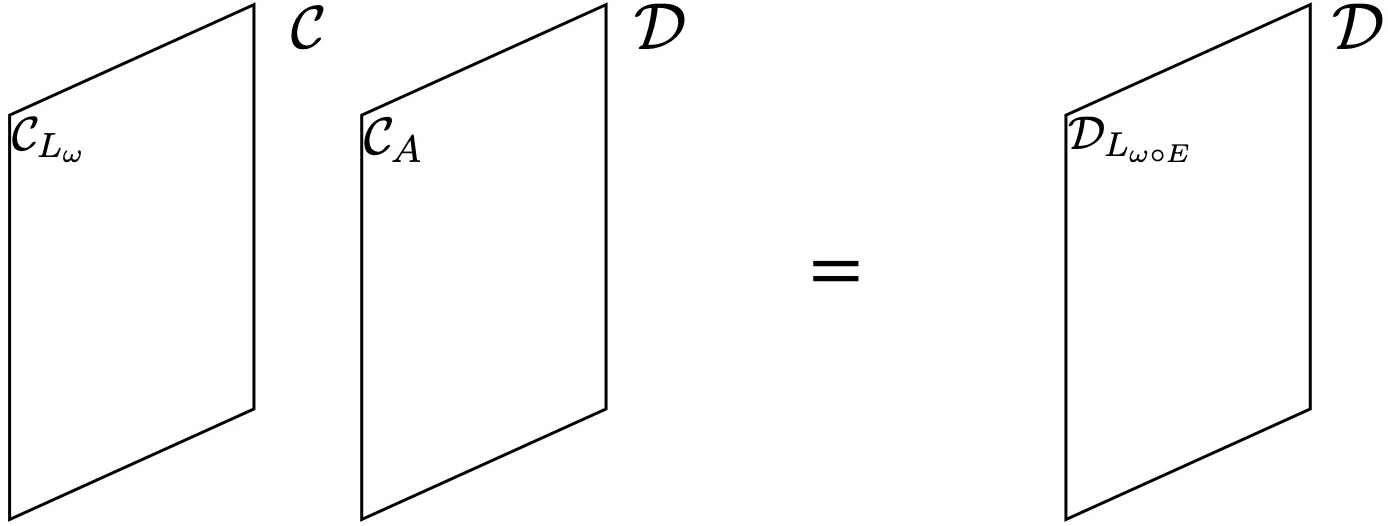}
\end{equation*}
where $L_{\omega \circ E}$ is generally a direct sum of Lagrangian algebras in $\mc{D} \cong \mathrm{DHR}(\mf{A})$ such that $\mc{D}_{L_{\omega \circ E}} \cong \mc{C}_{L_{\omega}} \boxtimes_{\mc{C}} \mc{C}_{A}$; the bimodule property~\eqref{eq:bimodule-property} is clearly satisfied. For the anyon condensation associated with the extension from $\mf{M}$ to $\mf{A}$, or equivalently, the condensable algebra $A$, there is an entropic order parameter; its value for each extension $\omega_{r}$ of the topological state $\omega$ is defined as the relative entropy
\begin{equation}
    S(\omega_{r} | \omega \circ E) = \mathrm{Tr}\left[ \rho_{\omega_{r}} \left( \log{\rho_{\omega_{r}}} - \log{\rho_{\omega \circ E}} \right) \right]
\end{equation}
between $\omega_{r}$ and $\omega \circ E$, where the density matrices are represented on the space $\mathrm{End}_{\mc{D}_{L_{\omega \circ E}}}(\boldsymbol{1})$, i.e., that spanned by the genuinely local operators in $\mf{A}$. The diagrammatics we developed throughout immediately allows one to find an upper bound on this entropic order parameter for pure states $\omega$ on $\mf{M}$. To this end, according to our discussions in subsection~\ref{subsec:connecting}, the aforementioned space can be divided into subspaces labelled by simple bulk lines $a \in \mathrm{Irr}(\mc{C})$~\emph{that can end topologically on both the boundary $\mc{C}_{L_{\omega}}$ and the interface $\mc{C}_{A}$}:
\begin{equation*}
\includegraphics[width=0.5\textwidth]{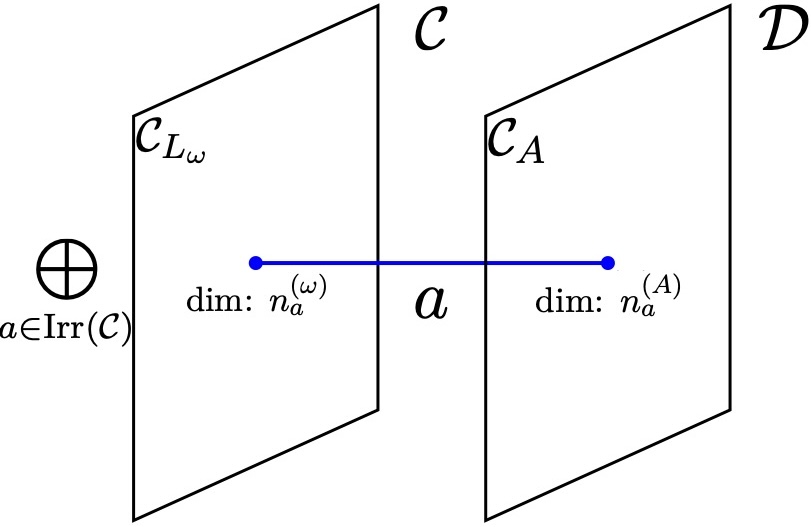}
\end{equation*}
For the condensable algebra~\eqref{eq:condensable-algebra}, the junction space of the line $a$ ending on the interface $\mc{C}_{A}$ is $n^{(A)}_{a}$-dimensional. Similarly, we express the Lagrangian algebra $L_{\omega}$ as
\begin{equation}
    L_{\omega} = \bigoplus_{a \in \mathrm{Irr}(\mc{C})} n^{(\omega)}_{a} a,
\end{equation}
which means that the junction space of $a$ on the boundary $\mc{C}_{L_{\omega}}$ is $n^{(\omega)}_{a}$-dimensional. The dimension of the space spanned by the genuinely local operators in $\mf{A}$ is then
\begin{equation}
\label{eq:dim-genuinely-local-operators}
    D = \sum_{a \in \mathrm{Irr}(\mc{C})} n^{(\omega)}_{a} n^{(A)}_{a}.
\end{equation}
On this space, there are $D$ pure extensions of the state $\omega$, i.e., one has $\omega_{r}$ with $r = 1, \ldots, D$. The density matrix $\rho_{\omega_{r}}$ is simply a rank-1 projection with the $r$-th diagonal entry equal to $1$ and all other entries $0$. From our discussions above, the identity line in the multi-fusion category $\mathcal{D}_{L_{\omega \circ E}}$ is nothing but $L_{\omega} \otimes A$ as an object in $\mathcal{C}$; $\omega \circ E$ is a mixed state with a weight for its projection on each minimal idempotent, again labelled by $r = 1, \ldots, D$, in $\mathrm{Hom}_{\mathcal{C}}(\boldsymbol{1}, L_{\omega} \otimes A)$. The bimodule property of $E$ implies that the weight is given by the coefficient in front of the unit operator in this idempotent; it is proportional to the trace of the corresponding projector, which, in turn, equals the quantum dimension of the image under this projection~\cite{fuchs2002,etingof2015}. Explicitly, denoting the image of $L_{\omega} \otimes A$ under the $r$-th projection as $X_{r}$ with quantum dimension $d_{r}$, we have
\begin{equation}
    L_{\omega} \otimes A = \bigoplus_{r=1}^{D} X_{r}, \quad \rho_{\omega \circ E} = \frac{1}{d_{L_{\omega}}d_{A}} \mathrm{diag} \left( d_{1}, \ldots, d_{D} \right),
\end{equation}
and hence
\begin{equation}
\label{eq:entropic-order-parameter-quantum-dimension}
    S(\omega_{r} | \omega \circ E) = \log{\frac{d_{L_{\omega}}d_{A}}{d_{r}}}.
\end{equation}
Note that $d_{A} = \sum_{a \in \mathrm{Irr}(\mc{C})} n^{(A)}_{a} d_{a}$ is the quantum dimension of the condensate $A$, and $d_{L_{\omega}}d_{A} = \sum_{r=1}^{D} d_{r}$. The bound on~\eqref{eq:entropic-order-parameter-quantum-dimension} now follows from a simple diagrammatic observation. Viewed from the boundary $\mathcal{C}_{L_{\omega}}$, the $r$-th component of the invisible line $L_{\omega} \otimes A$ is a boundary line, and $X_{r}$ collects precisely the bulk lines that can end topologically on it. The projection onto the $r$-th component is itself a genuinely local operator, which can be inserted on the trivial line in $\mathcal{C}_{L_{\omega}}$; namely, every component contains a copy of the trivial boundary line. Since the bulk lines that can end on the latter form the Lagrangian algebra $L_{\omega}$, one concludes that $L_{\omega}$ is contained in each $X_{r}$, and hence $d_{r} \geq d_{L_{\omega}}$, with equality if and only if $X_{r} \cong L_{\omega}$. We are thereby led to the bound\footnote{We note in passing that a constant term of entanglement entropy quantifying the `long-range entanglement' in a (2+1)d topological order, called the~\emph{topological entanglement entropy}~\cite{kitaev2006b,levin2006} $\gamma$, is directly related to the total quantum dimension of the MTC $\mc{C}$ describing this theory, $\gamma_{\mc{C}} = \frac{1}{2}\log{\mathrm{dim}(\mc{C})}$. The relation~\eqref{eq:relation-quantum-dimensions} then implies that $\gamma_{\mc{C}} - \gamma_{\mc{D}} = \log{d_{A}}$.}
\begin{equation}
\label{eq:entropic-order-parameter-bound}
    S(\omega_{r} | \omega \circ E) \leq \log{d_{A}}, \quad r = 1, \dots, D;
\end{equation}
when the bound is saturated, $A$ is a subobject of $L_{\omega}$.

Before entering the examples in the next section, we note that the situation is particularly simple when $\mathcal{C}$ is the quantum double of an Abelian finite group. In this case, the simple topological lines $a$ ending on both the boundary $L_{\omega}$ and the interface $A$ form a subgroup $K$ of the double with $|K| = D$, and $\mathrm{Hom}_{\mathcal{C}}(\boldsymbol{1}, L_{\omega} \otimes A)$ is simply the $K$-group algebra [there is no non-trivial multiplicity for any $a$ in either $L_{\omega}$ or $A$, because of the fact $d_{a} = 1$ and~\eqref{eq:decomposition-coefficients-constraint} in appendix~\ref{sec:condensable-algebras}]. For each $k \in K$, we can choose $v_{k}: \boldsymbol{1} \rightarrow k \otimes \bar{k}$ to be normalised so that $v_{k}v_{k^{\prime}} = v_{k+k^{\prime}}$ (`+' denotes the group multiplication in $K$), and it is readily verified that\footnote{See ref.~\cite{zhang2026} for a similar derivation, where the symmetry-breaking vacua play the role of idempotents.}
\begin{equation}
    P_{r} = \frac{1}{D} \sum_{k \in K} \chi_{r}(k)^{*} v_{k}
\end{equation}
are the minimal idempotents. Here, $\chi_{r}$ is the character function for the irreducible representation (irrep) labelled by $r$ \footnote{Recall that all irreps of an Abelian group are one-dimensional, which generate the Pontryagin dual that is isomorphic to the group itself.}. For any $r$, the coefficient in front of the unit $v_{0}$ is $\chi_{r}(0)^{*}/D = 1/D$, and
\begin{equation}
\label{eq:entropic-order-parameter-abelian}
    S(\omega_{r} | \omega \circ E) = \log{D}.
\end{equation}
In words, the value of entropic order parameter in this special case simply equals the logarithm of $D$, and is independent of $r$.

\section{Examples of entropic order parameter}
\label{sec:examples}

In this section, we compute the entropic order parameter for anyon condensations in several concrete examples and verify that the bound~\eqref{eq:entropic-order-parameter-bound} is satisfied.

\subsection{Toric code}

One of the simplest examples of a non-trivial TQFT is the toric code topological order. The latter is nothing but the $\mathbb{Z}_{2}$ Dijkgraaf-Witten (DW) gauge theory~\cite{dijkgraaf1990}, for which the MTC $\mc{C} = \mc{Z}(\mathrm{Rep}(\mathbb{Z}_{2}))$ is the quantum double of $\mathbb{Z}_{2} = \{ e, \eta \}$ ($e$ denotes the identity element and $\eta^{2} = e$) with total quantum dimension $\mathrm{dim}(\mc{C}) = 4$. In the quantum double of an Abelian finite group, each simple object is labelled by a pair
\begin{equation}  (\text{group element},~\text{irrep of the group}).
\end{equation}
$\mathbb{Z}_{2}$ has only two one-dimensional irreps, the trivial representation and the sign representation, denoted as $+$ and $-$, respectively. Explicitly, these bulk topological lines read
\begin{equation}
    \boldsymbol{1} = (e,+), \quad \mathbf{e} = (e, -), \quad \mathbf{m} = (\eta, +), \quad \mathbf{e} \otimes \mathbf{m} = (\eta, -),
\end{equation}
where $\mathbf{e}$ and $\mathbf{m}$ are bosonic, and their composition $\mathbf{e} \otimes \mathbf{m}$ is fermionic. In addition to the trivial algebra object $\boldsymbol{1}$ that condenses nothing (in which case the entropic order parameter vanishes for any $\omega$), there are two condensable algebras:
\begin{equation}
    A_{\mathbf{e}} = \boldsymbol{1} \oplus \mathbf{e}, \quad A_{\mathbf{m}} = \boldsymbol{1} \oplus \mathbf{m}.
\end{equation}
Both algebras are Lagrangian and lead to fully confining condensations, as their quantum dimensions are equal to $2 = \mathrm{dim}(\mathcal{C})^{1/2}$. On the other hand, the pure topological states $\omega$ are also labelled by these algebras. The entropic order parameters are given by~\eqref{eq:entropic-order-parameter-abelian}, where the dimensions $D$ defined in~\eqref{eq:dim-genuinely-local-operators} are computed and listed in the table below:
\begin{table}[H]
\centering
\begin{tabular}{|c|c|c|}
\hline
\diagbox{$L_{\omega}$}{$D$}{$A$} & \makecell{$A_{\mathbf{e}}$\\$d=2$} & \makecell{$A_{\mathbf{m}}$\\$d=2$} \\ \hline
$A_{\mathbf{e}}$ & \underline{2} & 1 \\ \hline
$A_{\mathbf{m}}$ & 1 & \underline{2} \\ \hline
\end{tabular}
\end{table}
\noindent
In this table and the one in the next subsection, the largest value of $D$ for a given condensation is underlined; in this case, it agrees with the quantum dimension of the corresponding condensable algebra.

\subsection{\texorpdfstring{$\mc{Z}(\mathrm{Rep}(\mathbb{Z}_{4}))$}{TEXT}
}

The $\mathbb{Z}_{4}$-DW theory, with MTC $\mc{C} = \mc{Z}(\mathrm{Rep}(\mathbb{Z}_{4}))$ the quantum double of $\mathbb{Z}_{4} = \{ e, \eta, \eta^{2}, \eta^{3}\}$ ($\eta^{4} = e$), is a generalisation of the toric code that is slightly more complicated. The four irreps of $\mathbb{Z}_{4}$ can again be labelled by the phase representing $\eta$: $1, \mathrm{i}, -1, -\mathrm{i}$. There are then 16 simple objects, or bulk topological lines, in $\mc{C}$:
\begin{equation}   \mathbf{e}^{i}\mathbf{m}^{j} = (\eta^{j}, \mathrm{i}^{i}), \quad i, j = 0, 1, 2, 3,
\end{equation}
where
\begin{equation}
    \mathbf{e} \equiv (e, \mathrm{i}) \quad \text{and} \quad \mathbf{m} \equiv (\eta, 1).
\end{equation}
There are 8 bosons, $\boldsymbol{1}, \mathbf{e}, \mathbf{e}^{2}, \mathbf{e}^{3}, \mathbf{m}, \mathbf{m}^{2}, \mathbf{m}^{3}, \mathbf{e}^{2}\mathbf{m}^{2}$, out of which 6 non-trivial condensable algebras can be composed:
\begin{align}
    &A_{1} = \boldsymbol{1} \oplus \mathbf{e}^{2}, \quad A_{2} = \boldsymbol{1} \oplus \mathbf{m}^{2}, \quad A_{3} = \boldsymbol{1} \oplus \mathbf{e}^{2}\mathbf{m}^{2}, \nonumber \\
    &A_{4} = \boldsymbol{1} \oplus \mathbf{e} \oplus \mathbf{e}^{2} \oplus \mathbf{e}^{3},~A_{5} = \boldsymbol{1} \oplus \mathbf{m} \oplus \mathbf{m}^{2} \oplus \mathbf{m}^{3},~A_{6} = \boldsymbol{1} \oplus \mathbf{e}^{2} \oplus \mathbf{m}^{2} \oplus \mathbf{e}^{2}\mathbf{m}^{2},
\end{align}
among which $A_{4}, A_{5}, A_{6}$ are Lagrangian algebras with quantum dimension $4 = \mathrm{dim}(\mathcal{C})^{1/2}$; they are the $L_{\omega}$ labelling the pure topological states $\omega$. The values of $D$ are computed and listed in the table below:
\begin{table}[H]
\centering
\begin{tabular}{|c|c|c|c|c|c|c|}
\hline
\diagbox{$L_{\omega}$}{$D$}{$A$} & \makecell{$A_{1}$\\$d=2$\\\text{toric code}} & \makecell{$A_{2}$\\$d=2$\\\text{toric code}} &
\makecell{$A_{3}$\\$d=2$\\\text{double semion}} & \makecell{$A_{4}$\\$d=4$} &
\makecell{$A_{5}$\\$d=4$} & \makecell{$A_{6}$\\$d=4$} \\ \hline
$A_{4}$ & \underline{2} & 1 & 1 & \underline{4} & 1 & 2 \\ \hline
$A_{5}$ & 1 & \underline{2} & 1 & 1 & \underline{4} & 2 \\ \hline
$A_{6}$ & \underline{2} & \underline{2} & \underline{2} & 2 & 2 & \underline{4} \\ \hline
\end{tabular}
\end{table}
\noindent
In the table, the post-condensation theories $\mc{C}_{A}^{\mathrm{loc}}$ for the non-Lagrangian condensable algebras $A$ are also indicated; we refer the reader to the book~\cite{simon2023} and papers~\cite{bais2009,eliens2014,neupert2016,burnell2018} for more relevant details. Again, the entropic order parameters are given by~\eqref{eq:entropic-order-parameter-abelian}, and the bound~\eqref{eq:entropic-order-parameter-bound} is saturated when $A$ is a subobject of $L_{\omega}$. We remark also that there is an obvious symmetry in the $D$-values for both $\mc{Z}(\mathrm{Rep}(\mathbb{Z}_{2}))$ and $\mc{Z}(\mathrm{Rep}(\mathbb{Z}_{4}))$ originating from the so-called electric-magnetic duality $\mathbf{e} \leftrightarrow \mathbf{m}$.

\subsection{\texorpdfstring{$\mc{Z}(\mathrm{Rep}(S_{3}))$}{TEXT}
}

Let us consider the quantum double $\mc{C} = \mc{Z}(\mathrm{Rep}(S_{3}))$ of the simplest non-Abelian group $S_{3} = \{ e, a, a^{2}, b, ab, a^{2}b \}$, which is the permutation group of three objects. Here, $a$ with $a^{3} = e$ generates a cyclic permutation, and $b$ with $b^{2} = e$ is a swapping; $a$ and $b$ do not commute with each other, $ba = a^{2}b \neq ba^{2} = ab$.

In the quantum double of a non-Abelian finite group, each simple object is labelled by a pair
\begin{equation}
    (\text{conjugacy class},~\text{irrep of the centraliser of a representative element in this class})
\end{equation}
generalising that for an Abelian finite group, as in the latter each conjugacy class contains only one element, and its centraliser is simply the group itself. For $S_{3}$, the three conjugacy classes are
\begin{equation}
    [e] = \{ e \}, \quad [a] = \{ a, a^{2} \}, \quad [b] = \{ b, ab, a^{2}b \},
\end{equation}
and the corresponding centralisers read
\begin{equation}
    \mathsf{C}_{S_{3}}(e) = S_{3}, \quad \mathsf{C}_{S_{3}}(a) = \{ e, a, a^{2}\} \cong \mathbb{Z}_{3}, \quad \mathsf{C}_{S_{3}}(b) = \{ e, b \} \cong \mathbb{Z}_{2}.
\end{equation}
Following our convention in the last two examples, the irreps of $\mathbb{Z}_{2}$ and $\mathbb{Z}_{3}$ are denoted as $\{ +, - \}$ and $\{ 1, \varphi, \varphi^{2} \}$ with $\varphi = \mathrm{e}^{2\pi\mathrm{i}/3}$, respectively. The three irreps of $S_{3}$ are the trivial representation $\mathbf{1}_{+}$, the sign representation $\mathbf{1}_{-}$, and the two-dimensional representation $\mathbf{2}$. One finds by a direct counting that $\mathrm{dim}(\mc{C}) = 36$. The simple anyons $\boldsymbol{1} \equiv ([e], \mathbf{1}_{+}), ([e], \mathbf{1}_{-}), ([e], \mathbf{2}), ([a], 1), ([b], +)$ are bosonic, out of which the following non-trivial condensable algebras can be composed:
\begin{align}
    &A_{1} = \boldsymbol{1} \oplus ([e], \mathbf{1}_{-}), \quad A_{2} = \boldsymbol{1} \oplus ([e], \mathbf{2}), \quad A_{3} = \boldsymbol{1} \oplus ([a], 1), \nonumber \\
    &A_{4} = \boldsymbol{1} \oplus ([e],\boldsymbol{1}_{-}) \oplus 2([e],\boldsymbol{2}), \nonumber \\
    &A_{5} = \boldsymbol{1} \oplus ([e],\boldsymbol{2}) \oplus ([b],+), \nonumber \\ 
    &A_{6} = \boldsymbol{1} \oplus ([e],\boldsymbol{1}_{-}) \oplus 2([a],1), \nonumber \\
    &A_{7} = \boldsymbol{1} \oplus ([a],1) \oplus ([b],+).
\end{align}
Their quantum dimensions are $d_{A_{1}} = 2, d_{A_{2}} = d_{A_{3}} = 3, d_{A_{4}} = d_{A_{5}} = d_{A_{6}} = d_{A_{7}} = 6$; $A_{4}, A_{5}, A_{6}, A_{7}$ are Lagrangian algebras labelling the pure topological states $\omega$. The values of $D$ are computed and listed in the table below:
\begin{table}[H]
\centering
\begin{tabular}{|c|c|c|c|c|c|c|c|}
\hline
\diagbox{$L_{\omega}$}{$D$}{$A$} & \makecell{$A_{1}$\\$d=2$\\$\mc{Z}(\mathrm{Rep}(\mathbb{Z}_{3}))$} & \makecell{$A_{2}$\\$d=3$\\\text{toric code}} & 
\makecell{$A_{3}$\\$d=3$\\\text{toric code}} & \makecell{$A_{4}$\\$d=6$} & 
\makecell{$A_{5}$\\$d=6$} & 
\makecell{$A_{6}$\\$d=6$} & 
\makecell{$A_{7}$\\$d=6$} \\ \hline
$A_{4}$ & 2 & 3 & 1 & 6 & 3 & 2 & 1 \\ \hline
$A_{5}$ & 1 & 2 & 1 & 3 & 3 & 1 & 2 \\ \hline
$A_{6}$ & 2 & 1 & 3 & 2 & 1 & 6 & 3 \\ \hline
$A_{7}$ & 1 & 1 & 2 & 1 & 2 & 3 & 3 \\ \hline
\end{tabular}
\end{table}
\noindent
We have computed the projections $X_{r}$ of $L_{\omega} \otimes A$
for each pair $(A, L_{\omega})$, from the quantum dimensions of which the entropic order parameters follow immediately according to~\eqref{eq:entropic-order-parameter-quantum-dimension}:
\begin{subequations}
\begin{align}
(A_{1},A_{4}):~X_{1}=X_{2}=A_{4};~d_{1}=d_{2}=6;~S_{1}=S_{2}=\log{2}.
\end{align}
\begin{align}
(A_{1},A_{5}):~X_{1}=A_{4}\oplus X^{(b)};~d_{1}=12;~S_{1}=0.
\end{align}
\begin{align}
(A_{1},A_{6}):~X_{1}=X_{2}=A_{6};~d_{1}=d_{2}=6;~S_{1}=S_{2}=\log{2}.
\end{align}
\begin{align}
(A_{1},A_{7}):~X_{1}=A_{6}\oplus X^{(b)};~d_{1}=12;~S_{1}=0.
\end{align}
\end{subequations}
\begin{subequations}
\begin{align}
(A_{2},A_{4}):~X_{1}=X_{2}=X_{3}=A_{4};~d_{1}=d_{2}=d_{3}=6;~S_{1}=S_{2}=S_{3}=\log{3}.
\end{align}
\begin{align}
(A_{2},A_{5}):~X_{1}=A_{5},~X_{2}=A_{4}\oplus X^{(b)};~d_{1}=6,~d_{2}=12;~S_{1}=\log{3},~S_{2}=\log{\frac{3}{2}}.
\end{align}
\begin{align}
(A_{2},A_{6}):~X_{1}=A_{4}\oplus2X^{(a)};~d_{1}=18;~S_{1}=0.
\end{align}
\begin{align}
(A_{2},A_{7}):~X_{1}=A_{5}\oplus X^{(a)}\oplus X^{(b)};~d_{1}=18;~S_{1}=0.
\end{align}
\end{subequations}
\begin{subequations}
\begin{align}
(A_{3},A_{4}):~X_{1}=A_{4}\oplus2X^{(a)};~d_{1}=18;~S_{1}=0.
\end{align}
\begin{align}
(A_{3},A_{5}):~X_{1}=A_{5}\oplus X^{(a)}\oplus X^{(b)};~d_{1}=18;~S_{1}=0.
\end{align}
\begin{align}
(A_{3},A_{6}):~X_{1}=X_{2}=X_{3}=A_{6};~d_{1}=d_{2}=d_{3}=6;~S_{1}=S_{2}=S_{3}=\log{3}.
\end{align}
\begin{align}
(A_{3},A_{7}):~X_{1}=A_{7},~X_{2}=A_{6}\oplus X^{(b)};~d_{1}=6,~d_{2}=12;~S_{1}=\log{3},~S_{2}=\log{\frac{3}{2}}.
\end{align}
\end{subequations}
\begin{subequations}
\begin{align}
(A_{4},A_{4}):~X_{1}=\ldots=X_{6}=A_{4};~d_{1}=\ldots=d_{6}=6;~S_{1}=\ldots=S_{6}=\log{6}.
\end{align}
\begin{align}
(A_{4},A_{5}):~X_{1}=X_{2}=X_{3}=A_{4}\oplus X^{(b)};~d_{1}=d_{2}=d_{3}=12;~S_{1}=S_{2}=S_{3}=\log{3}.
\end{align}
\begin{align}
(A_{4},A_{6}):~X_{1}=X_{2}=A_{4}\oplus2X^{(a)};~d_{1}=d_{2}=18;~S_{1}=S_{2}=\log{2}.
\end{align}
\begin{align}
(A_{4},A_{7}):~X_{1} = A_{4}\oplus2X^{(a)}\oplus3X^{(b)};~d_{1}=36;~S_{1}=0.
\end{align}
\end{subequations}
\begin{subequations}
\begin{align}
(A_{5},A_{5}):~&X_{1}=X_{2}=A_{5},~X_{3} = A_{4}\oplus X^{(a)}\oplus2X^{(b)};~d_{1}=d_{2}=6,~d_{3}=24; \nonumber \\
&S_{1}=S_{2}=\log{6},~S_{3}=\log{\frac{3}{2}}.
\end{align}
\begin{align}
(A_{5},A_{6}):~X_{1} = A_{4}\oplus2X^{(a)}\oplus3X^{(b)};~d_{1}=36;~S_{1}=0.
\end{align}
\begin{align}
(A_{5},A_{7}):~X_{1}=X_{2}=A_{5}\oplus X^{(a)}\oplus X^{(b)};~d_{1}=d_{2}=18;~S_{1}=S_{2}=\log{2}.
\end{align}
\end{subequations}
\begin{subequations}
\begin{align}
(A_{6},A_{6}):~X_{1}=\ldots=X_{6}=A_{6};~d_{1}=\ldots=d_{6}=6;~S_{1}=\ldots=S_{6}=\log{6}.
\end{align}
\begin{align}
(A_{6},A_{7}):~X_{1}=X_{2}=X_{3}=A_{6}\oplus X^{(b)};~d_{1}=d_{2}=d_{3}=12;~S_{1}=S_{2}=S_{3}=\log{3}.
\end{align}
\end{subequations}
\begin{align}
(A_{7},A_{7}):~&X_{1}=X_{2}=A_{7},~X_{3}=A_{1} \oplus ([e],\mathbf{2}) \oplus ([a],1) \oplus X^{(a)} \oplus 2X^{(b)}; \nonumber \\
&d_{1}=d_{2}=6,~d_{3}=24;~S_{1}=S_{2}=\log{6},~S_{3}=\log{\frac{3}{2}}.
\end{align}
The situations $(A_{5},A_{4})$, $(A_{6},A_{4})$, $(A_{6},A_{5})$, $(A_{7},A_{4})$, $(A_{7},A_{5})$, $(A_{7},A_{6})$ are the same as $(A_{4},A_{5})$, $(A_{4},A_{6})$, $(A_{5},A_{6})$, $(A_{4},A_{7})$, $(A_{5},A_{7})$, $(A_{6},A_{7})$, respectively, and are omitted. For notational simplicity, we have used the abbreviations $X^{(a)} \equiv ([a],1) \oplus ([a],\varphi) \oplus ([a],\varphi^{2})$, $X^{(b)} \equiv ([b],+) \oplus ([b],-)$, and $S_{r} \equiv S(\omega_{r} | \omega \circ E)$. The bound~\eqref{eq:entropic-order-parameter-bound} is saturated when (at least) one of the projections of $L_{\omega} \otimes A$ coincides with $L_{\omega}$.

\subsection{\texorpdfstring{$\mathrm{Fib} \boxtimes \overline{\mathrm{Fib}}$}{TEXT} and \texorpdfstring{$\mathrm{Ising} \boxtimes \overline{\mathrm{Ising}}$}{TEXT}}

Finally, let us look at two simple examples that go beyond the DW theory of a finite group, the `double Fibonacci' and `double Ising' TQFTs.

The Fibonacci fusion category $\mathrm{Fib}$ has only two simple objects, $\boldsymbol{1}$ and $\tau$, with the fusion rule $\tau \otimes \tau = \boldsymbol{1} \oplus \tau$; the non-integral quantum dimension of $\tau$ is the golden ratio, $d_{\tau} = (\sqrt{5}+1)/2$. The `double Fibonacci' TQFT is the product theory $\mc{C} = \mathrm{Fib} \boxtimes \overline{\mathrm{Fib}}$, where $\overline{\mathrm{Fib}}$ is the orientation reversal of $\mathrm{Fib}$ with the same objects but conjugated twist factors. The simple objects in $\mc{C}$ are denoted as $(\boldsymbol{1},\boldsymbol{1}), (\boldsymbol{1},\tau), (\tau,\boldsymbol{1}), (\tau,\tau)$, among which $(\boldsymbol{1},\boldsymbol{1})$ and $(\tau,\tau)$ are bosons. There is only one non-trivial condensable algebra, which is also Lagrangian:
\begin{equation}
    L = (\boldsymbol{1},\boldsymbol{1}) \oplus (\tau,\tau).
\end{equation}
$L$ itself also defines the only pure topological state $\omega$. One has $D=2$, $L \otimes L = X_{1} \oplus X_{2}$ with $X_{1} = L$ and $X_{2} = L \oplus (\boldsymbol{1},\tau) \oplus (\tau,\boldsymbol{1}) \oplus (\tau,\tau)$. As $d_{1} = 1 + d_{\tau}^{2}$ and $d_{2} = (1 + d_{\tau}^{2})d_{\tau}^{2}$,~\eqref{eq:entropic-order-parameter-quantum-dimension} yields $S(\omega_{1} | \omega \circ E) = \log{(1+d_{\tau}^{2})} = \log{d_{L}}$ and $S(\omega_{2} | \omega \circ E) = \log{(1 + d_{\tau}^{-2})}$.

Likewise, the MTC of the `double Ising' TQFT is $\mc{C} = \mathrm{Ising} \boxtimes \overline{\mathrm{Ising}}$. The well-known Ising fusion category has three simple objects $\boldsymbol{1}, \psi, \sigma$, with fusion rules $\psi \otimes \psi = \boldsymbol{1}, \psi \otimes \sigma = \sigma, \sigma \otimes \sigma = \boldsymbol{1} \oplus \psi$ and quantum dimensions $d_{\psi} = 1, d_{\sigma} = \sqrt{2}$. In $\mc{C}$, there are three bosons, $(\boldsymbol{1}, \boldsymbol{1}), (\psi, \psi), (\sigma, \sigma)$, out of which only two non-trivial condensable algebras can be composed:
\begin{equation}
    A = (\boldsymbol{1}, \boldsymbol{1}) \oplus (\psi, \psi), \quad L = (\boldsymbol{1}, \boldsymbol{1}) \oplus (\psi, \psi) \oplus (\sigma, \sigma).
\end{equation}
$A$ has quantum dimension $d_{A} = 2$, for which the condensation leads to the toric code topological order, whilst $L$ with $d_{L} = 4$ is Lagrangian and defines the only pure topological state $\omega$. The corresponding projections, their quantum dimensions, and the values of entropic order parameters are as follows:
\begin{align}
    (A,L):~&D=2;~X_{1} = X_{2} = L;~d_{1} = d_{2} = 4; \nonumber \\
    &S(\omega_{1} | \omega \circ E_{A}) = S(\omega_{2} | \omega \circ E_{A}) = \log{2} = \log{d_{A}}.
\end{align}
\begin{align}
    (L,L):~&D=3;~X_{1} = X_{2} = L,~X_{3} = L + (\boldsymbol{1}, \psi) + (\psi, \boldsymbol{1}) + (\sigma, \sigma); d_{1} = d_{2} = 4,~d_{3} = 8; \nonumber \\
    &S(\omega_{1} | \omega \circ E_{L}) = S(\omega_{2} | \omega \circ E_{L}) = \log{4} = \log{d_{L}},~S(\omega_{3} | \omega \circ E_{L}) = \log{2}.
\end{align}

\section*{Acknowledgements}
\addcontentsline{toc}{section}{Acknowledgements}

The author would like to thank Javier Molina-Vilaplana and Germ\'an Sierra for valuable discussions on related topics. He acknowledges support by an appointment to the Young Scientist Training (YST) Program at Asia Pacific Center for Theoretical Physics (APCTP) through the Science and Technology Promotion Fund and Lottery Fund of the Korean Government, and support by the Korean Local Governments - Gyeongsangbuk-do Province and Pohang City. The TikZ code for some of the figures/illustrations was written with the assistance of Gemini 1.5 Pro (Google) and GPT-5.6 Luna (OpenAI); the author has independently reviewed, verified, and takes full responsibility for all visual materials in this manuscript.

\appendix
\section{Condensable algebras}
\label{sec:condensable-algebras}

As mentioned, a condensable algebra is a~\emph{connected commutative symmetric special Frobenius algebra} object. For the sake of completeness, we present here the definition of these modifiers, and collect some constraints satisfied by the expansion coefficients of a condensable algebra in terms of the simple objects. We refer the reader to, e.g., refs.~\cite{fuchs2002,kong2014,bischoff2015} for more details.

An~\emph{algebra} in a tensor category $\mc{C}$ is an object $A \in \mc{C}$ equipped with two morphisms, the multiplication $\mu: A \otimes A \rightarrow A$ and the unit $u: \boldsymbol{1} \rightarrow A$ ($\boldsymbol{1}$ is the identity for the monoidal product $\otimes$) subject to associativity and left and right unit axioms, which are represented diagrammatically as follows (reading from bottom to top):
\begin{equation}
\begin{tikzpicture}[
    baseline=-0.5ex,
    scale=0.618,
    line width=1.2pt
]

\coordinate (L0) at (0,2.0);       
\coordinate (L1) at (-0.9,0.7);    

\draw (L0) -- (0,3.5)
    node[above=2pt] {$A$};

\draw (L0) -- (L1);

\draw (L1) -- (-1.7,-0.5)
    node[pos=0.9,below left=0pt] {$A$};

\draw (L1) -- (-0.25,-0.5)
    node[pos=0.9,below
    =0pt] {$A$};

\draw (L0) -- (1.5,-0.5)
    node[pos=0.9,below=2pt] {$A$};

\fill (L0) circle (2.5pt);
\node[below=3pt] at (L0) {$\mu$};

\fill (L1) circle (2.5pt);
\node[below=3pt] at (L1) {$\mu$};

\node at (2.25,1.35) {$=$};

\begin{scope}[xshift=4.5cm]

\coordinate (R0) at (0,2.0);       
\coordinate (R1) at (0.9,0.7);     

\draw (R0) -- (0,3.5)
    node[above=2pt] {$A$};

\draw (R0) -- (R1);

\draw (R1) -- (1.7,-0.5)
    node[pos=0.9,below right=0pt] {$A$};

\draw (R1) -- (0.25,-0.5)
    node[pos=0.9,below=0pt] {$A$};

\draw (R0) -- (-1.5,-0.5)
    node[pos=0.9,below=2pt] {$A$};

\fill (R0) circle (2.5pt);
\node[below=3pt] at (R0) {$\mu$};

\fill (R1) circle (2.5pt);
\node[below=3pt] at (R1) {$\mu$};

\end{scope}

\end{tikzpicture}, \qquad
\begin{tikzpicture}[
    baseline=-5.0ex,
    scale=0.618,
    line width=1.2pt
]

\coordinate (X0) at (-3.2,0.75);
\coordinate (X1) at (-4.1,-1.5);

\draw (X0) -- (-3.2,2.5)
    node[above=2pt] {$A$};
\draw (X0) -- (-3.2,-1.5);

\draw (X0) -- (X1);

\node[below=0pt] at (-3.2,-1.5) {$A$};

\fill (X0) circle (2.5pt);
\node[right=0pt] at (X0) {$\mu$};

\fill (X1) circle (2.5pt);
\node[below=3pt] at (X1) {$u$};

\node at (-1.5,0) {$=$};

\draw (0,-1.5) -- (0,2.5)
    node[above=2pt] {$A$};
\node[below=0pt] at (0,-1.5) {$A$};

\node at (1.5,0) {$=$};

\coordinate (Y0) at (3.2,0.75);
\coordinate (Y1) at (4.1,-1.5);

\draw (Y0) -- (3.2,2.5)
    node[above=2pt] {$A$};
\draw (Y0) -- (3.2,-1.5);

\draw (Y0) --(Y1);

\node[below=0pt] at (3.2,-1.5) {$A$};

\fill (Y0) circle (2.5pt);
\node[left=0pt] at (Y0) {$\mu$};

\fill (Y1) circle (2.5pt);
\node[below=3pt] at (Y1) {$u$};

\end{tikzpicture}.
\end{equation}
In addition to the algebra structure, a Frobenius algebra $A$ also admits two morphisms dual to the multiplication and the unit, i.e., the co-multiplication $\Delta: A \rightarrow A \otimes A$ and the co-unit $\varepsilon: A \rightarrow \boldsymbol{1}$ subject to co-associativity and co-unit axioms,
\begin{equation}
\begin{tikzpicture}[
    baseline=-11.5ex,
    scale=0.618,
    line width=1.2pt
]

\coordinate (L0) at (0,-2.0);
\coordinate (L1) at (-0.9,-0.7);

\draw (L0) -- (0,-3.5)
    node[below=2pt] {$A$};

\draw (L0) -- (L1);

\draw (L1) -- (-1.7,0.5)
    node[pos=1,above=0.5pt] {$A$};

\draw (L1) -- (-0.25,0.5)
    node[pos=1,above=0.5pt] {$A$};

\draw (L0) -- (1.5,0.5)
    node[pos=1,above=0.5pt] {$A$};

\fill (L0) circle (2.5pt);
\node[left=3pt] at (L0) {$\Delta$};

\fill (L1) circle (2.5pt);
\node[left=3pt] at (L1) {$\Delta$};

\node at (2.25,-1.35) {$=$};

\begin{scope}[xshift=4.5cm]

\coordinate (R0) at (0,-2.0);
\coordinate (R1) at (0.9,-0.7);

\draw (R0) -- (0,-3.5)
    node[below=2pt] {$A$};

\draw (R0) -- (R1);

\draw (R1) -- (1.7,0.5)
    node[pos=1,above=0.5pt] {$A$};

\draw (R1) -- (0.25,0.5)
    node[pos=1,above=0.5pt] {$A$};

\draw (R0) -- (-1.5,0.5)
    node[pos=1,above=0.5pt] {$A$};

\fill (R0) circle (2.5pt);
\node[right=3pt] at (R0) {$\Delta$};

\fill (R1) circle (2.5pt);
\node[right=3pt] at (R1) {$\Delta$};

\end{scope}

\end{tikzpicture}, \qquad
\begin{tikzpicture}[
    baseline=-7.5ex,
    scale=0.618,
    line width=1.2pt
]

\coordinate (X0) at (-3.2,-0.75);
\coordinate (X1) at (-4.1,1.5);

\draw (X0) -- (-3.2,-2.5)
    node[below=2pt] {$A$};
\draw (X0) -- (-3.2,1.5);

\node[above=2pt] at (-3.2,1.5) {$A$};

\draw (X0) -- (X1);

\fill (X0) circle (2.5pt);
\node[right=0pt] at (X0) {$\Delta$};

\fill (X1) circle (2.5pt);
\node[above=3pt] at (X1) {$\varepsilon$};

\node at (-1.5,-0.5) {$=$};

\draw (0,-2.5) -- (0,1.5);
\node[below=2pt] at (0,-2.5) {$A$};
\node[above=2pt] at (0,1.5) {$A$};

\node at (1.5,-0.5) {$=$};

\coordinate (Y0) at (3.2,-0.75);
\coordinate (Y1) at (4.1,1.5);

\draw (Y0) -- (3.2,-2.5)
    node[below=2pt] {$A$};
\draw (Y0) -- (3.2,1.5);

\node[above=2pt] at (3.2,1.5) {$A$};

\draw (Y0) -- (Y1);

\fill (Y0) circle (2.5pt);
\node[left=0pt] at (Y0) {$\Delta$};

\fill (Y1) circle (2.5pt);
\node[above=3pt] at (Y1) {$\varepsilon$};

\end{tikzpicture},
\end{equation}
which endows $A$ with the structure of a~\emph{co-algebra}. Furthermore, $\mu$ and $\Delta$ are required to satisfy the following~\emph{Frobenius property}:
\begin{equation}
\begin{tikzpicture}[baseline={(current bounding box.center)}, line width=1.2pt]
    
    \coordinate (L1) at (0.5, 0.5);
    \coordinate (L2) at (1.5, 1.5);
    
    \draw (0, 2) node[above] {$A$} -- (0, 1) to[out=-90, in=180] (L1);
    \draw (L2) to[out=0, in=90] (2, 1) -- (2, 0) node[below] {$A$};
    \draw (L1) to[out=0, in=180] (L2);
    \draw (L1) -- (0.5, 0) node[below] {$A$};
    \draw (L2) -- (1.5, 2) node[above] {$A$};
    
    \filldraw (L1) circle (1.2pt) node[above=2pt] {$\Delta$};
    \filldraw (L2) circle (1.2pt) node[below=2pt] {$\mu$};

    \node at (2.75, 1) {\Large $=$};

    \coordinate (M1) at (4.5, 1.25);
    \coordinate (M2) at (4.5, 0.75);
    
    \draw (3.7, 2) node[above] {$A$} -- (3.7, 1.5) to[out=-90, in=180] (M1);
    \draw (5.3, 2) node[above] {$A$} -- (5.3, 1.5) to[out=-90, in=0] (M1);
    \draw (M1) -- (M2);
    \draw (M2) to[out=180, in=90] (3.7, 0.5) -- (3.7, 0) node[below] {$A$};
    \draw (M2) to[out=0, in=90] (5.3, 0.5) -- (5.3, 0) node[below] {$A$};
    
    \filldraw (M1) circle (1.2pt) node[above=2pt] {$\Delta$};
    \filldraw (M2) circle (1.2pt) node[below=2pt] {$\mu$};

    \node at (6.25, 1) {\Large $=$};

    \coordinate (R1) at (7.5, 1.5);
    \coordinate (R2) at (8.5, 0.5);
    
    \draw (R1) to[out=180, in=90] (7, 1) -- (7, 0) node[below] {$A$};
    \draw (9, 2) node[above] {$A$} -- (9, 1) to[out=-90, in=0] (R2);
    \draw (R1) to[out=0, in=180] (R2);
    \draw (R1) -- (7.5, 2) node[above] {$A$};
    \draw (R2) -- (8.5, 0) node[below] {$A$};
    
    \filldraw (R1) circle (1.2pt) node[below=2pt] {$\mu$};
    \filldraw (R2) circle (1.2pt) node[above=2pt] {$\Delta$};
\end{tikzpicture}.
\end{equation}
$A$, as an algebra and a co-algebra at the same time, is called~\emph{special} if the mutually dual morphisms satisfy the condition\footnote{Different conventions for the coefficients on the right hand side of the two expressions in~\eqref{eq:specialness} are found in the literature. For example, one can also set both coefficients to be $\sqrt{d_{A}}$. It can be proven, however, that the product of these two coefficients equals the constant $d_{A}$~\cite{fuchs2002}.}
\begin{equation}
\label{eq:specialness}
\begin{tikzpicture}[baseline=-0.5ex, line width=1.2pt]

    \draw (0,-0.8) -- (0,0.8);
    \fill (0,0.8) circle (1.8pt) node[above=2pt] {$\varepsilon$};
    \fill (0,-0.8) circle (1.8pt) node[below=2pt] {$u$};
\end{tikzpicture} = d_{A}, \qquad
\begin{tikzpicture}[baseline=-0.5ex, line width=1.2pt]

    \draw (0,0.5) arc (90:270:0.4 and 0.5);
    \draw (0,0.5) arc (90:-90:0.4 and 0.5);
    \draw (0,0.5) -- (0,1.1) node[above] {$A$};
    \draw (0,-0.5) -- (0,-1.1) node[below] {$A$};

    \fill (0,0.5) circle (1.8pt) node[below=0pt] {$\mu$};
    \fill (0,-0.5) circle (1.8pt) node[above=0pt] {$\Delta$};
\end{tikzpicture}
\quad = \quad
\begin{tikzpicture}[baseline=-0.5ex, line width=1.2pt]

    \draw (0,-1.1) node[below] {$A$} -- (0,1.1) node[above] {$A$};
\end{tikzpicture},
\end{equation}
where $d_{A}$ is the~\emph{quantum dimension} of $A$. 

Suppose $\mc{C}$ is a~\emph{sovereign} tensor category~\cite{etingof2015}. $A$ is~\emph{symmetric} if
\begin{equation}
\begin{tikzpicture}[baseline={(current bounding box.center)}, line width=1.2pt]

    \coordinate (L1) at (0.5, 0.5);
    \coordinate (L2) at (1.5, 1.5);
    
    \draw (0, 2) node[above] {$\bar{A}$} -- (0, 1) to[out=-90, in=180] (L1);
    \draw (L2) to[out=0, in=90] (2, 1) -- (2, 0) node[below] {$A$};
    \draw (L1) to[out=0, in=180] (L2);
    \draw (L2) -- (1.5, 2);
    
    \filldraw (L2) circle (1.2pt) node[below=2pt] {$\mu$};
    \filldraw (1.5, 2) circle (1.2pt) node[above=2pt] {$\varepsilon$};

    \node at (4.5, 1) {\text{(as morphism: $A \rightarrow \bar{A}$)}~\Large $=$};

    \coordinate (R1) at (7.5, 1.5);
    \coordinate (R2) at (8.5, 0.5);
    
    \draw (R1) to[out=180, in=90] (7, 1) -- (7, 0) node[below] {$A$};
    \draw (9, 2) node[above] {$\bar{A}$} -- (9, 1) to[out=-90, in=0] (R2);
    \draw (R1) to[out=0, in=180] (R2);
    \draw (R1) -- (7.5, 2);

    \filldraw (R1) circle (1.2pt) node[below=2pt] {$\mu$};
    \filldraw (7.5, 2) circle (1.2pt) node[above=2pt] {$\varepsilon$};
    
\end{tikzpicture}
\end{equation}
with $\bar{A}$ the orientation reversal of $A$.

Suppose $\mc{C}$ is a~\emph{braided} tensor category with braiding $c_{X,Y}: X \otimes Y \rightarrow Y \otimes X$. An algebra object $A \in \mc{C}$ is called~\emph{commutative} if
\begin{equation}
    \mu \circ c_{A,A} = \mu.
\end{equation}
Finally, an algebra object $A$ is~\emph{connected} if
\begin{equation}
    \mathrm{Hom}(\boldsymbol{1},A) \cong \mathbb{C}.
\end{equation}

Any condensable algebra in a unitary MTC $\mc{C}$ is expressed in terms of the simple objects as
\begin{equation}
\label{eq:condensable-algebra-append}
    A = \bigoplus_{a \in \mathrm{Irr}(\mc{C})} n_{a} a,
\end{equation}
which implies that the quantum dimension $d_{A} = \sum_{a \in \mathrm{Irr}(\mc{C})} n_{a} d_{a}$. It turns out that the coefficients $n_{a} \in \mathbb{Z}_{\geq 0}$ have to obey the following constraints~\cite{chatterjee2023}:
\begin{subequations}
\begin{equation}
    n_{\boldsymbol{1}} = 1, \quad n_{a} = n_{\bar{a}};
\end{equation}
\begin{equation}
\frac{\sum_{b}\mathbb{S}_{ab}n_{b}}{\sum_{b}\mathbb{S}_{\boldsymbol{1}b}n_{b}}~\text{is a cyclotomic integer for any $a$, $\mathbb{S}$ is the modular S-matrix of $\mc{C}$};
\end{equation}
\begin{equation}
\label{eq:decomposition-coefficients-constraint}
    n_{a} \leq \begin{cases}
\begin{array}{c}
d_{a}, \\
\lfloor d_{a} \rfloor - 1,
\end{array} & \begin{array}{c}
d_{a}~\text{is integral,} \\
\text{otherwise;}
\end{array} \end{cases}
\end{equation}
\begin{equation}
    n_{a}n_{b} \leq \sum_{c} \mathbb{N}_{ab}^{c}n_{c} - \delta_{a\bar{b}} (\lceil d_{a} \rceil - \lfloor d_{a} \rfloor),~\mathbb{N}~\text{is the matrix of fusion coefficients of $\mc{C}$};
\end{equation}
\begin{equation}
    n_{a} = \sum_{b} \mathbb{S}_{ab}n_{b} \quad \text{if $A$ is Lagrangian}.
\end{equation}
\end{subequations}

\printbibliography

\end{document}